\documentclass{aa}
\usepackage{txfonts}
\usepackage{natbib}
\usepackage{longtable}
\usepackage[pdftex]{graphicx}
\usepackage{multirow}
\usepackage[bookmarks,breaklinks]{hyperref}
   \hypersetup{
     colorlinks,
     citecolor=blue,
     filecolor=magenta,
     linkcolor=blue,
     urlcolor=cyan
}

\begin{document}

\title{Deconstructing blazars: A different scheme for jet kinematics in flat-spectrum AGN}

   \author{M. Karouzos\inst{1,2}
        \fnmsep\thanks{Member of the International Max Planck Research School (IMPRS) for Astronomy and Astrophysics at the Universities of Bonn and Cologne}\fnmsep\thanks{Current affiliation: Center for the Exploration of the Origin of the Universe, Seoul National University, Gwanak-gu, Seoul 151-742, Korea}
      \and S. Britzen\inst{1}
           \and A. Witzel\inst{1}
              \and J.A. Zensus\inst{1,3}
     \and A. Eckart\inst{3,1}}

   \institute{ Max-Planck-Institut f\"ur Radioastronomie, Auf dem H\"ugel 69, 53121 Bonn, Germany
   \and
   Reimar-L\"ust Fellow of the Max Planck Society\\
            \email{mkarouzos@astro.snu.ac.kr}
   \and
   I.Physikalisches Institut, Universit\"at zu K\"oln, Z\"ulpicher Str. 77, 50937 K\"oln, Germany
            }


   \date{Received / Accepted}

\abstract{Recent VLBI studies of the morphology and kinematics of individual BL Lac objects (S5 1803+784, PKS 0735+178, etc.) have revealed a new paradigm for the pc-scale jet kinematics of these sources. Unlike the apparent superluminal outward motions usually observed in blazars, most, if not all, jet components in these sources appear to be stationary with respect to the core, while exhibiting strong changes in their position angles. As a result, the jet ridge lines of these sources evolve substantially, at times forming a wide channel-flow.}
{We investigate the Caltech-Jodrell Bank flat-spectrum (CJF) sample of radio-loud active galaxies to study this new kinematic scenario for flat-spectrum AGN. Comparing BL Lac objects and quasars in the CJF, we look for differences with respect to the kinematics and morphology of their jet ridge lines. The large number of sources in the CJF sample, together with the excellent kinematic data available, allow us to perform a robust statistical analysis in that context.}
{We develop a number of tools that extract information about the apparent linear and angular evolution of the CJF jet ridge lines, as well as their morphology. In this way, we study both radial and non-radial apparent motions in the CJF jets. A statistical analysis of the extracted information allows us to test this new kinematic scenario and assess the relative importance of non-radial and radial motions in flat-spectrum AGN jets, compared to those of quasars. We also use these tools to check the kinematics for (multi-wavelength) variable AGN.}
{We find that approximately half of the sample shows appreciable apparent jet widths ($>10 degrees$), with BL Lac jet ridge lines showing significantly larger apparent widths than both quasars and radio galaxies. In addition, BL Lac jet ridge lines are found to change their apparent width more strongly. Finally, BL Lac jet ridge lines show the least apparent linear evolution, which translates to the smallest apparent expansion speeds for their components. We find compelling evidence supporting a substantially different kinematic scenario for flat-spectrum radio-AGN jets and in particular for BL Lac objects. In addition, we find that variability is closely related to the properties of a source's jet ridge line. Variable quasars are found to show ``BL Lac like" behavior, compared to their non-variable counterparts.}
{}


\authorrunning{Karouzos et al.}
\titlerunning{A different scheme for jet kinematics in BL Lacs}
   \maketitle

\section{Introduction}
\label{sec:intro}

Although observed in the minority of active galaxies ($\sim5-15\%$; \citealt{Kellermann1989}; \citealt{Padovani1993}; \citealt{Jiang2007}), extragalactic jets are some of the most pronounced morphological features in AGN research. Their presumably direct connection to the active core and the supermassive black hole (SMBH) residing there, makes them invaluable tools in the effort to characterize the properties and the underlying physics of activity in galaxies. VLBI observations enable the direct imaging of AGN jets and thus the study of their properties on parsec scales. One of the most prominent discoveries, related to jet kinematics, was that of apparent superluminal motion of jet components (e.g., \citealt{Whitney1971}; \citealt{Pearson1981}), a combination of relativistic expansion speeds (close to the speed of light) and the projected geometry onto the plane of the sky (e.g., \citealt{Rees1966}).

Jet kinematics, as studied through the investigation of distinct components, is usually explained in terms of the shock-in-jet model (e.g., \citealt{Marscher1985}), where the observed jet knots are manifestations of shocks propagating down the jet at relativistic speeds. Beaming and projection effects regulate the observed properties of the jets. There has been continuous effort to distinguish whether the different types of active galaxies (e.g., quasars, BL Lacs, Fanarrof-Riley, etc.) are just a result of orientation effects, or if additionally these objects have intrinsically different properties. The current paradigm is that indeed different jet properties can be attributed to geometrical effects combined with factors such as the black hole mass or the accretion rate. For example, \citet{Ghisellini1993b} for a sample of 39 superluminal sources find no appreciable difference between the distribution of Doppler factors between BL Lacs and flat-spectrum radio-quasars (FSRQs), with RGs showing smaller values. Some indications to the contrary also exist (e.g., \citealt{Gabuzda1995}; \citealt{Gabuzda2000}). Analysis of statistically important samples (large number of sources and/or stringent selection criteria) of active galaxies have been of fundamental importance to this end (e.g., \citealt{Ghisellini1993}; \citealt{Vermeulen1994}; \citealt{Vermeulen1995}; \citealt{Taylor1996}; \citealt{Hough2002}; \citealt{Lister2005}; \citealt{Britzen2007a}).

Although it is crucial to pursue a statistical approach to the open problem of AGN jet kinematics, the study of individual sources is indispensable. It has helped to elucidate particular mechanisms or effects that might get smoothed out by the poor temporal or spatial resolution usually available in big statistical samples. Such a case of a detailed study of an individual object is that of S5 1803+784.

\subsection{S5 1803+784: A case study}
\label{sec:1803}

S5 1803+784 is an active galaxy at a moderate redshift of z=0.68 (\citealt{Hewitt1989}). It has been classified as a BL Lac object. Being a member of the complete S5 sample (\citealt{Witzel1987}) it has been extensively studied in the radio, at different wavelengths and with different instruments (see \citealt{Britzen2010} for a detailed recounting of the source's radio observations history). Typical for this class of objects, 1803+784 has been observed to be variable in the radio and the optical on both long and short-timescales (e.g., \citealt{Wagner1995}; \citealt{Heidt1996}; \citealt{Nesci2002}; \citealt{Fan2007}).

\citet{Britzen2010}, by (re-)analyzing more than 90 epochs of global VLBI and VLBA data, reveal a new kinematic scheme for 1803+784. All components in the inner part of the jet (up to 12 mas) appear to remain stationary with respect to the core. This behavior is seen at all frequencies studied by the authors (1.6 - 15 GHz). In contrast to this, the components show strong changes in their position angles, implying a prevailing movement perpendicular to the jet axis.

\citet{Britzen2010} also studied the jet ridge line morphology and evolution of S5 1803+784. A jet ridge line at a given epoch is defined as the line that linearly connects all component positions at that epoch. The authors find that the jet ridge line changes in an almost periodic manner, starting resembling a straight line, evolving into a sinusoid-like pattern, and finally returning to its original linear pattern, although slightly displaced from its original position. A period of $\sim 8.5$ years is calculated for the evolution of the jet ridge line.

Finally, the authors find that the jet changes its apparent width (in a range between a few and a few tens of degrees) in an almost periodic way with a timescale similar to the one found from the evolution of the jet ridge line. All of the above properties support a new kinematic scheme for 1803+784, where components follow oscillatory-like trajectories, with their movement predominantly happening perpendicular to the jet axis rather than along it. Moreover, the jet appears at times to form a wide channel of flow, while changing its width considerably across time.

\subsection{Motivation}
S5 1803+784 is one of several BL objects exhibiting the behavior described above. 0716+714 has been shown to behave in a similar way (\citealt{Britzen2009}), with most of its components being stationary with respect to the core while changing their position angles considerably. PKS 0735+178 is another example of a source with similar, but rather more complicated, kinematic properties (\citealt{Gomez2001}; \citealt{Agudo2006}; \citealt{Britzen2010b}). Under the unification scheme of active galaxies (e.g., \citealt{Antonucci1993}; \citealt{Urry1995}), both BL Lac objects and FSRQs are believed to be active galaxies for which the viewing angle to their jets is very small, leading to strong relativistic effects. Given the similar viewing angle distributions, in the last several years the general classification of blazar has often been used to describe members of either class, also in terms of their jet properties. However, this phenomenological unification of FSRQs and BL Lacs comes into question in light of the recent investigation of sources like S5 1803+784, PKS 0735+178 and 0716+714. Is the peculiar kinematic behavior seen in these objects revealed due to the unprecedented richness of the datasets available, and could therefore be relevant for all flat-spectrum radio-AGN, or are BL Lacs characterized by a genuinely different set of kinematic properties? 

This paper focuses on the relevance of this new kinematic scheme for flat-spectrum radio-AGN, while investigating the apparent divide between BL Lac and FSRQ jet kinematics. As valuable as single source studies are to an in-depth understanding of particular phenomena, there are a number of biases or unaccounted factors that alter and ultimately hinder a universal application of their results. In this context, we use the CJF sample to statistically investigate and assess the similarity, or divergence, of the kinematic and morphological properties between the two distinct sub-samples of FSRQs and BL Lac objects in the CJF. We want to test whether jet components of BL Lac objects indeed show slower apparent speeds with respect to their cores compared to FSRQs. Furthermore we are interested in the phenomenon exhibited in S5 1803+784 of an, at times, very wide jet, as well as a strong evolution of that width. For this investigation we use tools that extract information from the jet ridge line of the sources, instead of focusing on individual components. This allows for a investigation that is mostly independent of component modeling and cross-identification. 

The paper is organized as follows: in Sect. \ref{sec:1803} we introduce the new kinematic scheme for BL Lac objects, as shown for the case of the source 1803+784, and discuss the motivation of this work, in Sect. \ref{sec:cjf} we describe the CJF sample, in Sect. \ref{sec:data} we describe the data used, in Sect. \ref{sec:analysis} we present the analysis of our data and the results, and in Sect. \ref{sec:discussion} we discuss our results and give some conclusions. Throughout the paper, we assume the cosmological parameters $H_{0}=71$ $km s^{-1} Mpc ^{-1}$, $\Omega_{M}=0.27$, and $\Omega_{\Lambda}=0.73$ (from the first-year WMAP observations; \citealt{Spergel2003}).

\section{The CJF sample}
\label{sec:cjf}

\begin{figure}
\begin{center}
  \includegraphics[width=0.4\textwidth,angle=0]{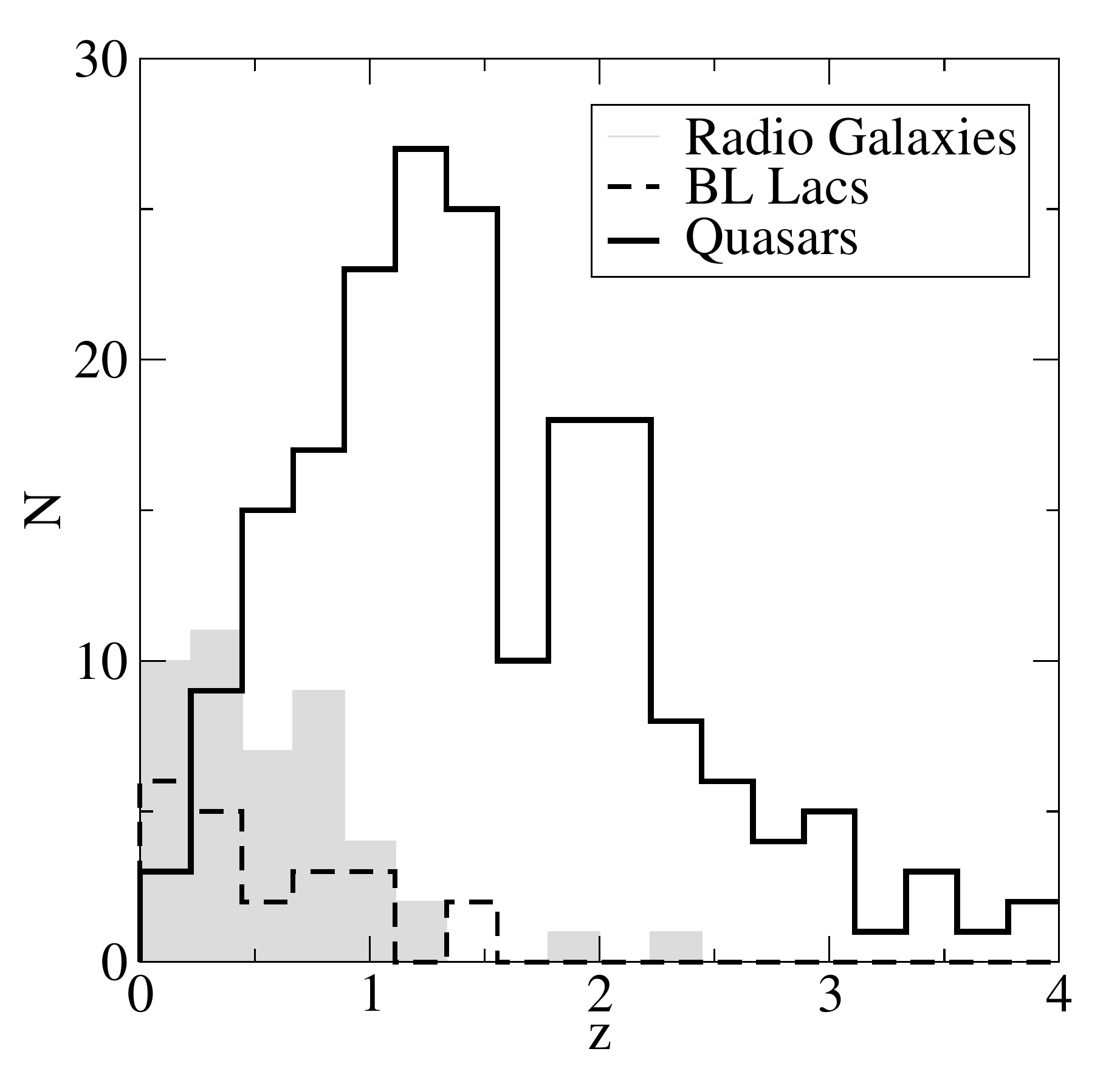}
  \caption{Redshift distribution of radio galaxies (grey blocks), BL Lacs (dashed line), and FSRQs (solid line) in the CJF sample.}
  \label{fig:redshift_histo}
\end{center}
\end{figure}

The CJF sample (\citealt{Taylor1996}) consists of 293 radio-loud active galaxies selected (see Table \ref{tab:cjfproperties}) from three different samples (for details, see \citealt{Britzen2007a}). The sources span a large redshift range (see Fig. \ref{fig:redshift_histo}), the farthest object being at a redshift $z=3.889$ (1745+624; \citealt{Hook1995}) and the closest at $z=0.0108$ (1146+596; \citealt{deVaucouleurs1991}). The average redshift of the sample is $z_{avg}=1.254$, $z_{BL Lac,avg}=0.546$, $z_{RG,avg}=0.554$, and $z_{FSRQ,avg}=1.489$ for BL Lacs, radio galaxies, and FSRQs, respectively. All the objects have been observed  with the VLBA and/or the global VLBI network. Each source has at least 3 epochs of observations (with a maximum of 5 epochs) and has been imaged and studied kinematically (\citealt{Britzen1999}; \citealt{Britzen2007a}; \citealt{Britzen2008}). The X-ray properties have been studied and correlated with their VLBI properties (\citealt{Britzen2007b}). The evolution of active galaxies, in the context of the merger-driven evolution scheme (e.g., \citealt{Hopkins2006}), has also been investigated with the help of the CJF, identifying candidate CJF sources in different evolutionary stages, including new binary black hole candidates (\citealt{Karouzos2010}).

\begin{table}[htp]
\caption{CJF sample and its properties.}
\label{tab:cjfproperties}
\begin{tabular}{l c}
\hline\hline
\textbf{Frequency(MHz)}      &  4850 \\
\textbf{Flux lower limit @5GHz}    &  350mJy\\
\textbf{Spectral Index}      &  $\alpha_{1400}^{4850}\geq -0.5$ \\
\textbf{Declination}         &  $\delta\geq 35^{\circ}$  \\
\textbf{Galactic latitude}   &  $|b|\geq 10^{\circ}$ \\
\textbf{\# Quasars}           &  198 \\
\textbf{\# BL Lac }           &  32 \\
\textbf{\# Radio Galaxies}    &  52 \\
\textbf{\# Unclassified}      &  11  \\
\textbf{\# Total}             &  293 \\
\hline
\end{tabular}
\end{table}

\subsection{Radio emission}

The CJF is a flux-limited radio-selected sample of flat-spectrum radio-loud AGN. The sample was originally created to study, among other things, the kinematics of pc-scale jets and apparent superluminal motion (see \citealt{Taylor1996} for more details).

The CJF sample (Table \ref{tab:cjfproperties}) has been most extensively studied in the radio regime (e.g., \citealt{Taylor1996}; \citealt{Pearson1998}; \citealt{Britzen1999}; \citealt{Vermeulen2003}; \citealt{Pollack2003}; \citealt{Lowe2007}; \citealt{Britzen2007a}; \citealt{Britzen2008}). \citet{Britzen2008} developed a localized method for calculating the bending of the jet associated with individual components. The maximum of the distribution of local angles is at zero degrees, although a substantial fraction shows some bending ($0-40$ degrees). A few sources exhibit sharp bends of the order of $>50$ degrees (see Fig. 13 in \citealt{Britzen2008}).

Although the CJF sample consists mostly of core-dominated AGN, presumably highly beamed sources, the kinematical study of the sample identifies a large number of sources with stationary, subluminal, or, at best, mildly superluminal outward velocities (e.g., see Fig. 15 in \citealt{Britzen2008}). Combined with a number of sources with inwardly moving components (e.g., 0600+422, 1751+441, 1543+517, \citealt{Britzen2007a}), these sources do not fit into the regular paradigm of outward, superluminaly moving components in blazar jets. One explanation of these peculiar kinematic behaviors is that of a precessing, or helical jet (e.g., \citealt{Conway1993}) possibly as a result of a SMBH binary system. Other interpretations include trailing shocks produced in the wake of a single perturbation propagating down the jet (e.g., \citealt{Agudo2001}; but also see \citealt{Mimica2009}), or standing re-collimation shocks (e.g., \citealt{Gomez1995}).

\citet{Karouzos2010} compile a list of all the CJF sources that have been found to show long timescale variability in the radio (as well as in other wavelength regimes). This list comprises in total 40 CJF sources, 27 of which have been argued to show possibly periodical variability of their fluxes. The authors do not take into account intra-day variability.

\section{Data}
\label{sec:data}

The work presented here is heavily based on the kinematic analysis of the CJF sample (\citealt{Britzen2007a}; \citealt{Britzen2008}). An extensive observing campaign of all 293 CJF sources was undertaken using both the VLBA and the global VLBI array at 5 GHz (see \citealt{Britzen2007a} for details). We note that five CJF sources were initially excluded from any further analysis due to problematic observations (0256+424, 0344+405, 0424+670, 0945+664, 1545+497) and therefore are also omitted here. In total, 288 sources are considered and analyzed in the following sections. Of these, according to the optical classification from \citet{Britzen2007a}, 196 are classified as quasars, 49 as radio galaxies, 33 as BL Lac objects , and 10 are not classified.

Due to the scope of the CJF program, the identification and analysis of pc-scale jet component kinematics has focused on the part of the jet that is beamed towards us. For a number of sources, several components belonging to the counter-jet have been identified. However, for these sources cross-identification of the counter-jet components over epochs has not been carried out. For this reason, and given the nature of the analysis that we undertook (see below), we have excluded all counter-jet components in the following investigation. In a total number of 2468 components identified, 82 (3.32\%) counter-jet components have been identified and excluded from our analysis. \citet{Britzen2007a} report that on average radio galaxies have 3.6 components identified per jet, 2.7 components are identified per quasar jet, and 2.9 components per BL Lac jet. This reflects the relative difference of projected jet length for the different types of object classes. This shall be discussed more thoroughly in Sect. \ref{sec:discussion}.

In the following, the tools that we use for the analysis of the CJF jet ridge lines, in the context described in Sect. \ref{sec:1803}, are described. In short we used the following measures:
\begin{itemize}
\item Monotonicity Index, M.I.
\item Apparent Jet Width, dP
\item Apparent Jet Width Evolution, $\Delta P$
\item Apparent Jet Linear Evolution, $\Delta\ell$
\end{itemize}
We note that all the above measures, as their names imply, refer strictly to values projected onto the plane of the sky. Although it is possible to constrain, or in some cases have a specific estimate of the viewing angle of each source and therefore attempt to calculate intrinsic jet properties, this is outside the scope of this paper. For the following sections, we adopt the basic assumption as described by the AGN unification scheme (e.g., \citealt{Antonucci1993}, \citealt{Urry1995}) that BL Lacs and FSRQs are seen at the smallest viewing angles, while radio galaxies have their jet axis further away from our line of sight. As we are mainly interested in the comparison between FSRQs and BL Lacs, deprojection of the jet properties investigated here is not critical. For the following analysis, and for the sake of brevity, we shall drop the characterization of apparent for each of these values, although this will be implied throughout unless otherwise stated.

\begin{figure*}[ht]
\begin{center}
  \includegraphics[width=0.5\textwidth,angle=0]{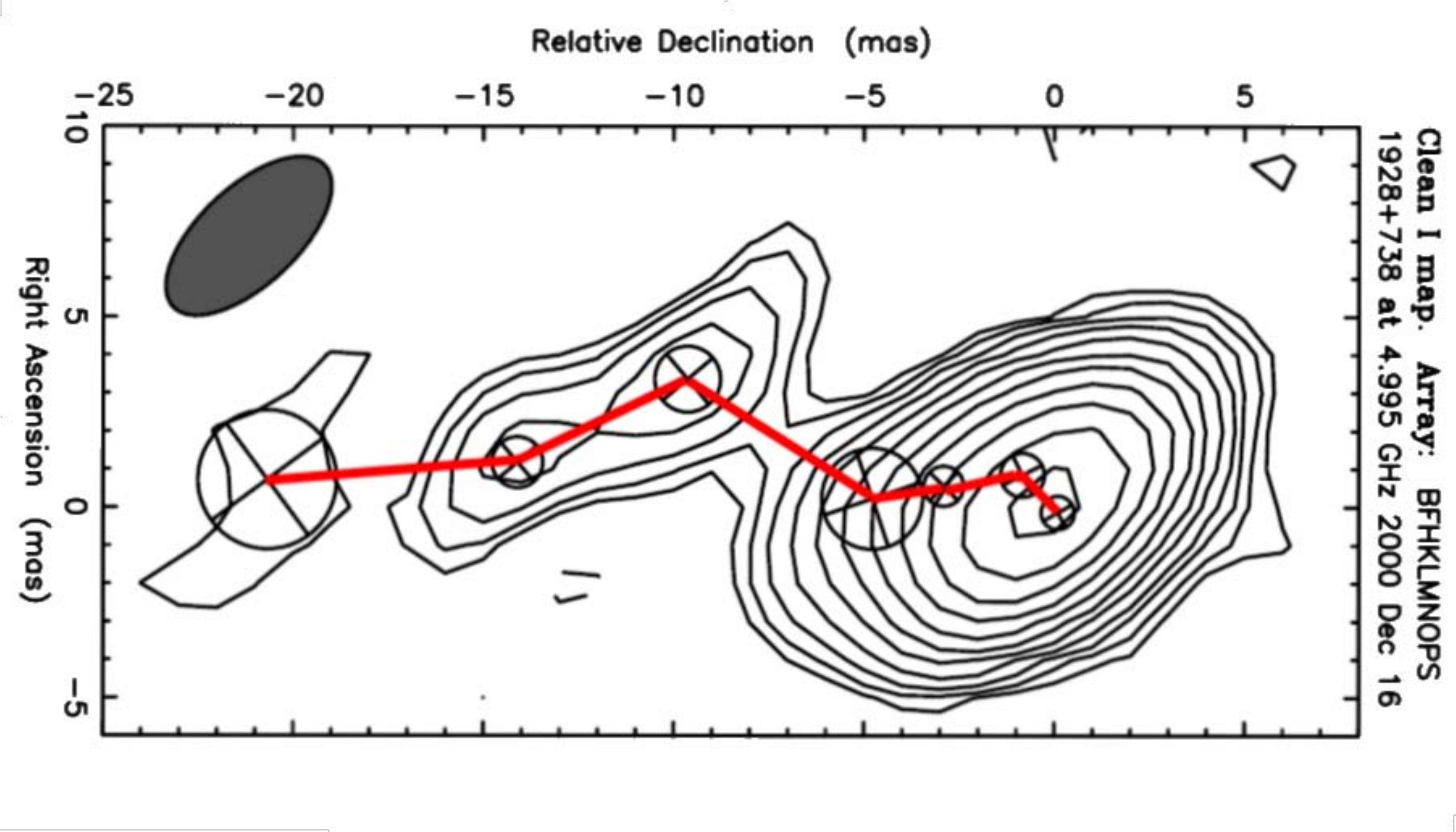}\hspace{40pt}
  \includegraphics[width=0.4\textwidth,angle=0]{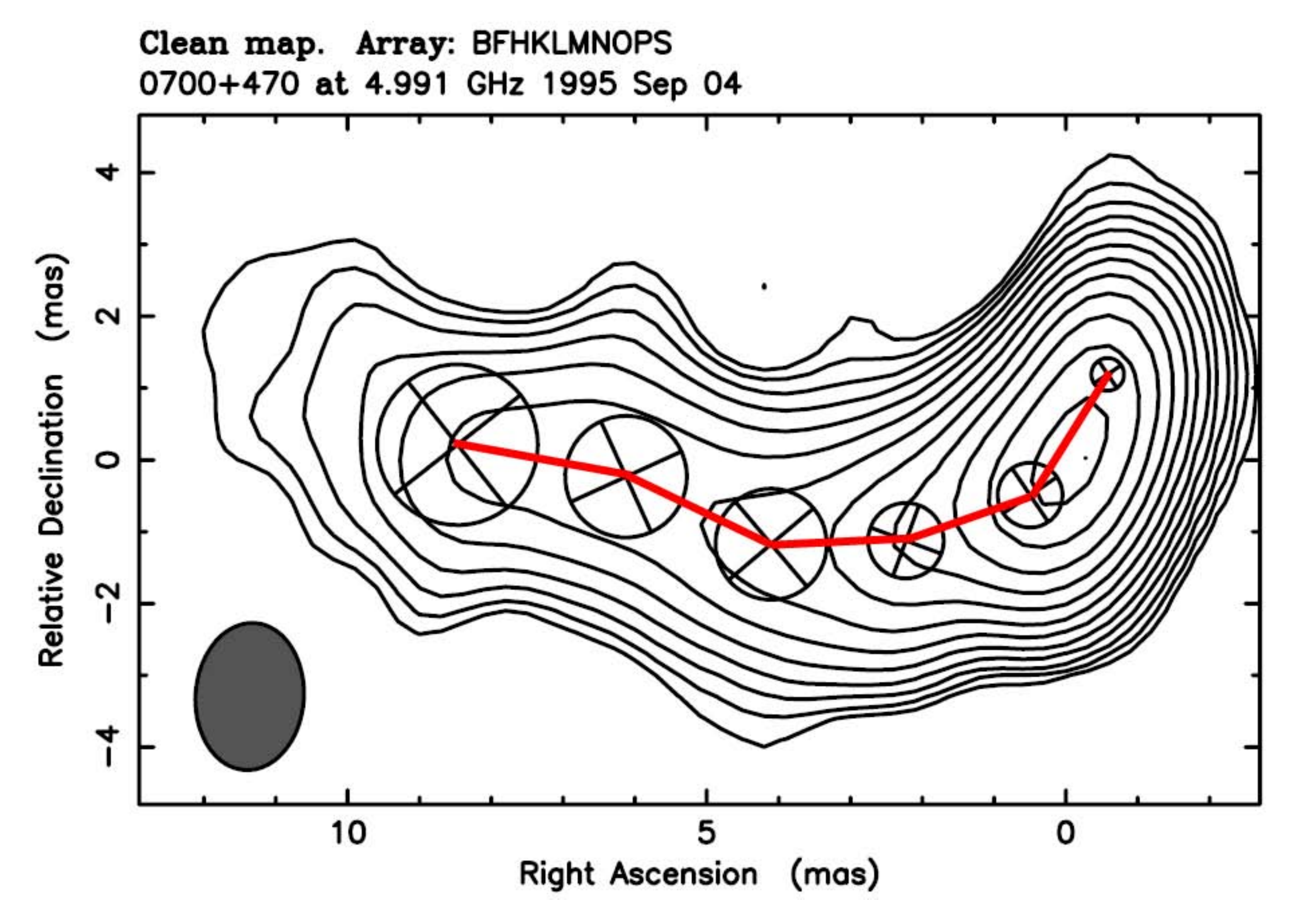}
  \caption[Jet Ridge Line M.I. Example]{Radio maps at 5GHz of 1928+738 (left) and 0700+470 (right), with jet ridge lines superimposed. Two examples of a sinusoid-like jet morphology (left) and a single-bent one (right). VLBI maps from \citet{Britzen2007a}.}
  \label{fig:ridge_lines_example}
\end{center}
\end{figure*}

\subsection{Monotonicity Index, M.I.}
\label{sec:rl.MI}

It is known that a large number of active galaxies exhibit bent or otherwise non-linear jet morphologies on various scales. Individual sources like S5 1803+784 and PKS 0735+178 (see references above), as well as others (e.g. 3C 345; \citealt{Lobanov1999}; B0605-085; \citealt{Kudryavtseva2010}), have been extensively studied to understand the origin of these bends. \citet{Britzen2008} calculate the ``bends" between successive components of the pc-scale jet for all CJF sources, finding large local variations of the position angles of the sources. In this context, we are interested in quantifying the bending of the whole jet ridge line. Moreover, we want to differentiate between a ``monotonically-bent'' jet, i.e., a jet that is bent in only one direction (see Fig. \ref{fig:ridge_lines_example}, right), as opposed to a more sinusoid-like morphology (similar to what is seen for some epochs of S5 1803+784, see Fig. \ref{fig:ridge_lines_example}, left). This is done by means of the monotonicity index, M.I..

We quantify a sinusoid-like morphology of a jet by identifying the local extrema in a given jet ridge line, in the core separation - position angle plane. For an epoch \emph{i}, a component \emph{m} exhibits a local extremum under the definition
$$\theta_{m;extr}:|\theta_{m}-\theta_{m\pm1}|\geq 10(d\theta_{m}+d\theta_{m\pm1}),$$
where $\theta_{m}$ and $d\theta_{m}$ denote the position angle of component and its uncertainty. $\theta_{m}$ is calculated as
$$\theta_{m}=\arctan\frac{X}{Y},$$
where X and Y are cartesian coordinates on the plane of the sky.
Having calculated the number of extrema for a given jet ridge line at an epoch i, we define the M.I. as
$$M.I.=\frac{number\;\;of\;\;extrema}{N-1},$$
where N is the total number of components at that given epoch. This is a crude calculation, but can give us a handle on how the bending of the jet behaves along the jet. For M.I. values close to one, the jet resembles more a sinusoid. An M.I. value close to zero reveals a monotonic, single-bend, jet morphology. We normalize for the number of components N to account for longer, or shorter, jets and to enable comparison between different sources. As an example, for the sources shown in Fig. \ref{fig:ridge_lines_example}, 1928+738 is found to have an M.I. value of 0.4, compared to 0700+470, that gives an M.I. value of 0.

As is the case for most of the tools described in this section, the value of M.I. depends heavily on the resolution of the observations. Given that the resolution is not the same across all epochs and sources, an uncertainty is introduced when comparing two M.I. values of two different sources. A final remark pertains to the definition of an extremum. We use a $10\sigma$\footnote{Here $\sigma$ is defined as the sum of the position angle errors for $\theta_{m}$ and $\theta_{m\pm1}$. The choice of this $\sigma$ reflects an original underestimation (in \citealt{Britzen2007a}) of the position angle errors during the component fitting.} value as the lower limit for flagging an extremum. A different level of significance (e.g., $5\sigma$) would result in different values of M.I.. The choice of the $10\sigma$ significance is a conservative approach to the identification of extrema in the jet ridge line. In the following, M.I. shall be used as more of a qualitative tool rather than a quantitative measure of the actual jet morphology.

\subsection{Apparent jet width, dP}
\begin{figure}[b]
\begin{center}
  \includegraphics[width=0.4\textwidth,angle=0]{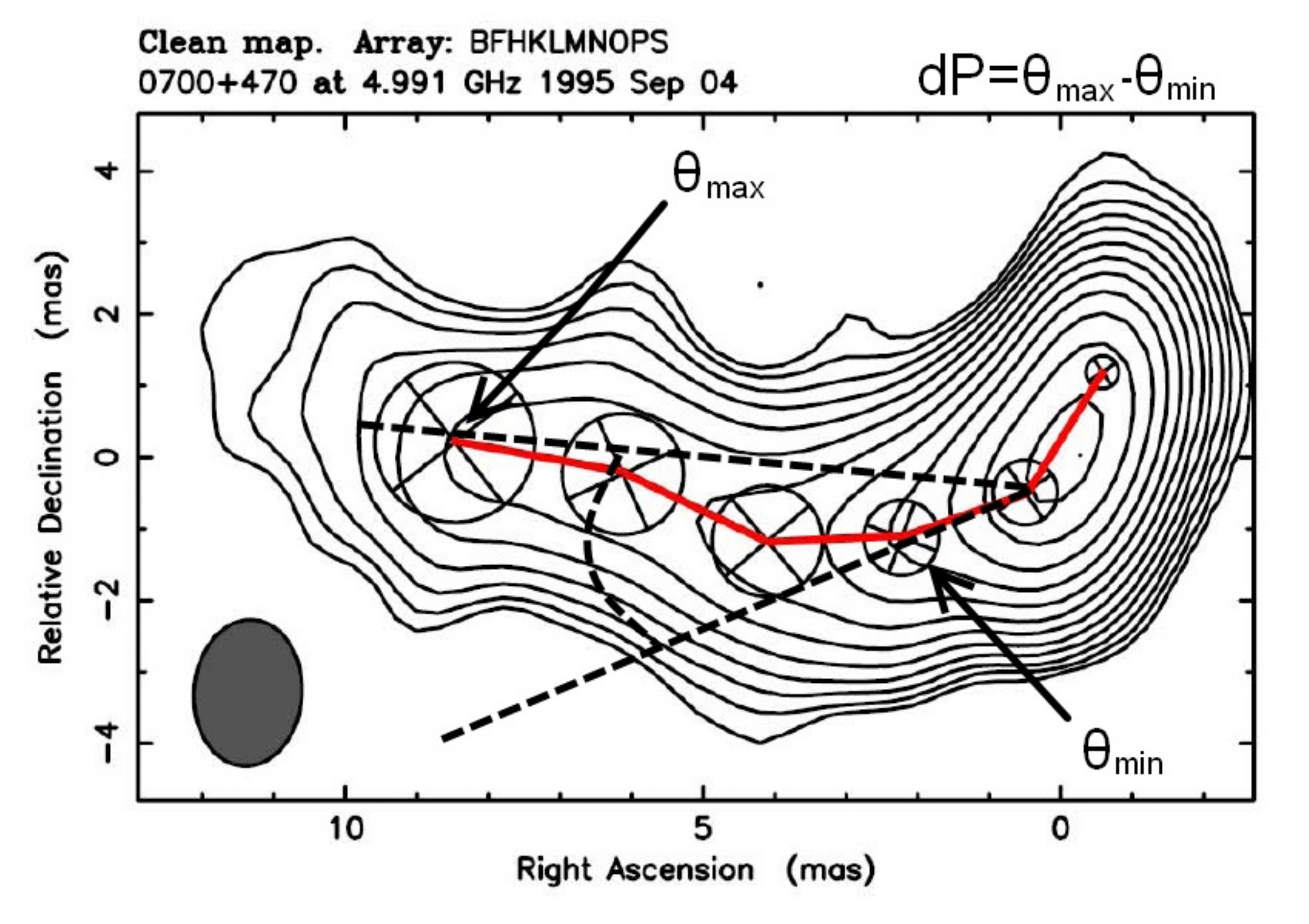}
  \caption{Radio map of 0700+470 at 5GHz, exhibiting the calculation of the jet ridge line width. The two arrows denote the components at maximum and minimum position angle, while the area in between is defined as the opening, or apparent width, of the jet ridge line. The VLBI core is found at (0,0) coordinates. VLBI map from \citet{Britzen2007a}}
  \label{fig:ridge_line_width_example}
\end{center}
\end{figure}

For an epoch \emph{i} and a jet consisting of N components characterized by their core separation and position angle $(r_{i},\theta_{i})$, we identify the components with the maximum and the minimum position angles. The apparent width of the jet dP, measured in degrees, is then calculated as (see Fig. \ref{fig:ridge_line_width_example}):
$$dP_{i}=\theta_{i}^{max}-\theta_{i}^{min},$$
while the error is calculated by the propagation of errors formula.
There is a number of factors that should be considered at this point. The most obvious drawback for the above definition is the non-localized nature of this measure. Using two different components, at different core separations, gives us only an approximate notion of the width of the flow. Higher resolution investigations might resolve the structure of the jet perpendicular to the jet axis and provide us with a more real estimate of the local width of the jet. Alternatively, one can calculate a localized value of the jet width by using the FWHM of the fitted Gaussian components (e.g., \citealt{Pushkarev2009}). In this case however, the width is heavily dependent on the beam size and therefore on the resolution. Instead, by using the definition above, we seek to quantify the opening of the jet flow and identify the effect of a channel-like jet, as seen in the case of S5 1803+784. In the context of the localization problem, we have implemented in our calculations a constraint in the maximum allowed distance between the two components identified with the minimum and maximum position angles. The value of this limit depends on the jet length and shall be discussed in Sect. \ref{sec:analysis}.

It is clear that in the case of a narrow but highly bent jet, dP would then measure the amplitude of the bending rather than the width of the flow itself. In the context of this study such an effect does not present a problem as we are, to first order, interested in the final result, i.e., an apparently wide jet ridge line opening, rather than whether this effect is due to a genuinely wide jet flow or rather a high-amplitude bending. 

\subsection{Apparent jet width evolution, $\Delta P$}
The jet width evolution $\Delta P$, measured in degrees per unit time, is measured between two successive epochs \emph{i} and \emph{(i-1)}. It is calculated as follows:
$$\Delta P=\frac{dP_{i}-dP_{i-1}}{T_{i}-T_{i-1}},$$
where $T_{i}$ denotes the time at epoch \emph{i}, measured in years. We can also define the maximum jet width evolution as:
$$\Delta P^{max}=max\{\Delta P_{1...i}\}.$$
This value is characteristic of each source and reflects the maximum potential width change of the jet flow for that source. We should note here that the jet width evolution does not reflect an angular speed (as implied by the units of degrees per unit time). Instead the per unit time reflects a normalization for time that ensures a comparison between sources that have been observed during different time spans. The currently available data are too sparse to allow us a calculation of a real angular speed for how the jet width (and orientation) changes with time.

For both the width dP and the width evolution $\Delta P$, the sensitivity and dynamic range of our observations play a regulating role. Given that we need at least two components to define the width, a dimming of one of these components across epochs can result to a false or altered value for both the calculated width and width evolution. A more detailed discussion of this can be found in Sect. \ref{sec:discussion}.

\subsection{Apparent jet linear evolution, $\Delta\ell$}
We calculate the linear evolution across all available epochs and for all components, ultimately acquiring a value reflecting the total linear displacement of the whole jet ridge line. We use plane-of-the-sky coordinates $(X_{i},Y_{i})$ to calculate the linear displacement of component \emph{m }between epochs \emph{i} and \emph{(i+1)}:
$$l^{m}_{i}=\sqrt{\Delta X^{2}+\Delta Y^{2}}.$$
To calculate the total displacement of the whole jet ridge line we then need to sum up over all components and all available observing epochs:
$$\ell=\sum_{i-1}\sum_{m}l^{m}_{(i)}.$$
We need to account for both the different time span of observations, as well as the different number of components. Therefore we define the jet linear evolution $\Delta\ell$, measured in parsecs per unit time and per component, as follows:
$$\Delta\ell=\frac{\sum_{i-1}\sum_{m}l^{m}_{(i)}}{{N(T_{i}-T_{1})}}=\frac{\ell}{NdT},$$
where N here is the total number of components used across all the epochs.
We need to underline a fundamental difference between the way the jet linear evolution, $\Delta\ell$, is calculated, compared to the measures described previously. For the calculation of the displacement of an individual component $\ell$ between two consecutive epochs, and consequently of $\Delta\ell$, the cross-identification of components across epochs is necessary. Therefore, for $\Delta\ell$, unlike the previously discussed jet ridge line characteristic values, the actual identification of components is important.

The reason for normalizing this value over the time span of observations is the same as for the jet width evolution. In this case however we also need to account for the number of components for each individual jet. We therefore divide the total displacement of the whole jet ridge line by the actual number of components used for the calculation (hence the number of components cross-identified across each pair of observation epochs).

$\Delta\ell$ essentially reflects the apparent speed distribution of all cross-identified components of the jet and therefore represents a value characteristic for the whole jet, rather than for any individual component. By summing up all components and epochs we trade temporal and positional resolution for a universal treatment of the entire jet. In this way we can test whether the kinematics of BL Lac objects is fundamentally different than that of FSRQs while averaging out localized properties of individual components. $\Delta\ell$ can be seen as a mean jet component speed, with the difference that it is acquired through averaging not only over all components, but additionally over all available epochs. Although it certainly reflects a measure of the outward motion in BL Lac jets, the calculation of $\Delta\ell$ is done in such a way, that the potential curvature of the components' trajectories is taken into account. This separates $\Delta\ell$ from a simple linear regression fit to the core-separation versus time diagrams usually employed to calculate outward velocities, making it sensitive to non-radial motions, that are otherwise missed.\\
\\
Lastly, given the large span of redshifts that the CJF sources cover, we must account for the different linear scales probed for sources at different distances from us. By comparing sources at different redshifts, one implicitly studies a different part of their jet. To counter this effect, we explicitly define a maximum apparent core separation limit, above which components are not included in our investigation. There are however additional arguments that indicated the necessity of such a limit, both technical as well as physical. It is known that AGN jets appear most curved closer to the core, with bending effects becoming dumped at larger distances (e.g., \citealt{Krichbaum1994}, \citealt{Britzen2000}). It should also be noted that at larger separation from the reference center of a map (the core), the uncertainties involved in the identification and fitting of jet components increase. By constraining our analysis in the inner-structure of the jets, we partly compensate for this effect. In this way we make certain that we probe the same linear scales for all CJF sources, albeit with different resolution. The latter effect can be counteracted by using redshift bins throughout. Additionally, through use of redshift bins, we can probe a possible evolution of the studied properties with redshift, if any. Given that the separation limit used is an apparent one, a level of uncertainty, dependent on the scatter of the viewing angle distribution within a certain object class, is expected.

\section{Analysis and results}
\label{sec:analysis}

As the apparent jet length (expressed by the maximum distance of the outermost jet component identified in each source, across all available epochs) and its morphology (as expressed by the M.I.) are used as a basis for the further analysis, we are going to briefly discuss these first.

\subsection{Apparent jet length and morphology}
We find BL Lacs to show shorter jets on parsec scales in median values, compared to quasars and radio galaxies ($17.3\pm2.7$ pc compared to $27.1\pm2.4$ pc and $43\pm7$ pc;$\sim3\sigma$ level), with their average values however not being statistically significantly different  ($34\pm5$ pc, compared to $31\pm4$ pc and $50\pm6$ pc, respectively; $<3\sigma$)\footnote{Throughout the paper we calculate median errors like $1.253\sigma/\sqrt(N)$, where $\sigma$ is the standard deviation of the mean and N is the number of values used. We assume that the quantities studied here follow a normal distribution.}. Having calculated the average and median apparent jet lengths for the CJF objects, we are able to define a maximum apparent core separation limit (for the reasons discussed in Sect. \ref{sec:data}). As we are interested in the behavior of BL Lacs, we use them as the basis of our decision. For 1803+784, peculiar kinematics of the components is observed for the inner-most part of its jet (up to 12 mas), with most robust effects out to 6 mas. For this source's redshift (z=0.68), this translates roughly to (82 pc) 41 pc. Similarly to 1803+784, 0735+178 and 0716+714 also show the most prominent evolution of their jet ridge line out to $\sim6$ mas (39.6 pc and 33 pc, for their respective redshifts). Given that most of the BL Lacs in our sample is at $z<1$, the length scale implied by the three sources mentioned above (the only objects whose jet ridge line has been investigated in detail) ensures that the inner-jet of sources at z$<1$ is covered. We adopt therefore an apparent core separation limit of 40 parsecs. All components at separations larger than 40 parsecs with respect to the core are not included in the following analysis. For a limit of 40 pc we probe the whole jet for 60\% of the CJF sources (72\% for BL Lacs, 60\% for radio galaxies, and 56\% for FSRQs), while for approximately 17\% of the CJF sources more than half of the jet is excluded from further analysis.

Adopting a core distance limit somewhat reduces the available number of components and therefore possibly increases the scatter in the following statistical analysis. Moreover, this limit implicitly gives a higher weight to more strongly beamed sources. This bias however applies equally to both BL Lacs and FSRQs, and should therefore not influence the comparison between the two classes. The higher number of components ``missed" for the FSRQs because of this limit is mainly due to their redshift distribution and therefore using redshift bins should minimize the bias to our results.

We investigate how, if at all, a jet resembles a sinusoid (as in the case of S5 1803+784). In order to do this we calculate the M.I. from the jet ridge lines of our sources. The M.I. is calculated for the whole jet (we do not apply the core separation limit of 40 pc), as we are interested in the morphology of the entire jet ridge line, rather than a localized property. Additionally, a large number of components is needed for a robust interpretation of the M.I., we therefore take into account only epochs with at least three components identified. Finally, we define the maximum M.I. value for each source across the available epochs. Fluctuations of the M.I. of a source between epochs are low.

\begin{figure}[h]
\begin{center}
  \includegraphics[width=0.5\textwidth,angle=0]{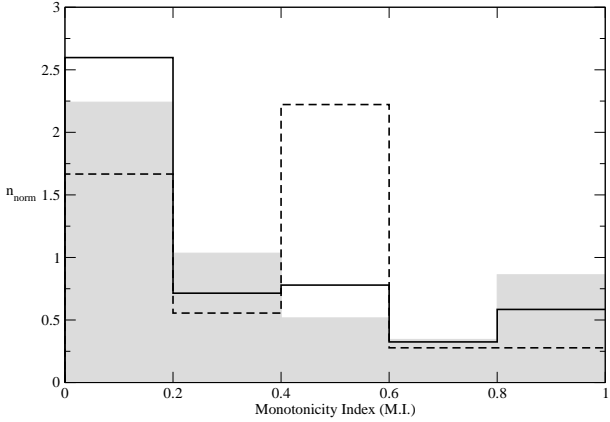}
  \caption{Monotonicity Index (M.I.) distribution for BL Lacs (dashed line), FSRQs (solid line), and RGs (grey blocks). The histogram has been normalized to area unity.}
  \label{fig:MI_types_histo}
\end{center}
\end{figure}

Under the above constraints, we find that there are in total 46 sources with M.I. $\geqslant0.5$. This translates to 37\% of the total number of sources used here. BL Lacs show more often jets with sinusoid-like morphologies, compared to both FSRQs and RGs. More than half of the FSRQs ($\sim 51\%$) have an M.I. value of 0, while two thirds ($\sim 66\%$) have an M.I. value lower than 0.5. For RGs the respective percentages are $\sim 41\%$ and $\sim 66\%$. In contrast to this, only one third of the BL Lacs ($\sim 33\%$) have M.I.=0 and less than half ($\sim 44\%$) of them have M.I.$<0.5$. Figure \ref{fig:MI_types_histo} shows the histogram of the M.I. for the different types of objects. BL Lacs show their maximum in the [0.4,0.6) bin. In contrast, both FSRQs and RGs have their maxima in the [0,0.2) bin. In Table \ref{tab:MI_types} the statistical properties of the different types of objects are given. It can be seen that the average M.I. values for BL Lacs and FSRQs agree within the $1\sigma$ errors. When considering the median values for the three classes of objects, the difference becomes more apparent, with BL Lacs and FSRQs differing at $>3\sigma$ level. It should be noted that a robust quantitative treatment of the M.I. is problematic, given the severe restrictions imposed by the small number of components per jet, as well as the limited temporal resolution of the observations used here.

\begin{table}[htb]
\begin{center}
\caption{Statistical properties of the monotonicity index (M.I.) distribution for FSRQs, BL Lacs, and RGs. Average and median are calculated for all sources with at least three components in their jet. Only the maximum M.I. values are considered for each source.}
\label{tab:MI_types}
\begin{tabular}{l | l l l}
\multicolumn{4}{l}{}\\
\multicolumn{4}{l}{\large{\textbf{Monotonicity Index (max)} }}\\
\hline
\textbf{Types}	&\textbf{FSRQ}	&\textbf{BL}	&\textbf{RG} \\
\hline
\#	            	 &77  		&18   	&29   			\\
\textbf{Average}&0.29		&0.36	&0.35 		          \\
\textbf{Error}	&0.03 		&0.06 	&0.06 			\\
\textbf{Median}	&0 			&0.500	&0.330 			\\
\textbf{Error}	&0.004		&0.014	&0.012			\\
\hline
\end{tabular}
\end{center}
\end{table}

In this somewhat qualitative context BL Lacs appear to more often show sinusoid-like curved jets, as seen for S5 1803+784 and 0716+714.

\subsection{Apparent jet width, P, statistics}
In Sect. \ref{sec:data}, we have already briefly discussed the concept of localized width and how this relates to the measure we are going to investigate here. In order to better understand this relation, we study how the distribution of the widths of the CJF jet ridge lines relates to the maximum distance between the two components used to calculate this width. For one epoch of the source S5 1803+784 and one epoch of the source 2200+420, a width of $\sim180$ degrees is found. This is probably due to a problem in the original component fitting and we therefore exclude these two epochs in this analysis.

We calculate non-zero jet widths for a total of 557 epochs (multiple epochs per source). Of these, 277 are larger than 10 degrees, while 125 are larger than 20 degrees. This translates to 22.4\% of the sample possessing jets with widths exceeding 20 degrees.

We calculated an average jet ridge line width for our sources. Without imposing any constraints on the maximum distance between the $\theta_{max}$ and $\theta_{min}$ components used, we find that BL Lacs show significantly wider jet ridge lines (average value of $19.3\pm1.8^\circ$), compared to FSRQs ($12.0\pm0.4\circ$; $4\sigma$ difference). RGs do not show significant differences in their mean width compared to BL Lacs. A Student's t-test gives a significance of $>$99.99\% that BL Lacs and FSRQs show different mean values. After imposing a constraint of 20 pc for the maximum distance between the $\theta_{max}$ and $\theta_{min}$ components, the BL Lacs still show statistically wider jet ridge lines. Low number statistics does not allow us to draw a robust conclusion for more localized widths (at $\sim5$ pc linear scales) of our CJF jet ridge lines.

We also investigated sources in the redshift bin [0,1]. Although the redshift span of the CJF goes out to almost 4, of the 32 BL Lacs included in the CJF, only 23 have redshift information. Of these, 20 are below redshift 1. Therefore for the following we will focus our investigation to sources up to redshift 1. In Table \ref{tab:jet_width_statistics_z_1} we show the characteristic statistical parameters for the CJF jet ridge line widths for all sources with available redshift, as well as sources in the redshift bin [0,1]. Given the above analysis concerning the distance constraints, in Table \ref{tab:jet_width_statistics_z_1} we give values also for the case where a 20 pc constraint for the maximum distance between the $\theta_{max}$ and $\theta_{min}$ components is assumed.

\begin{table*}
\begin{center}
\caption{Characteristic statistical values concerning the width of the jet ridge line of sources with measured redshifts. We give average and median values with uncertainties, maximum, and minimum values for FSRQs, BL Lacs, and RGs. The noted distance pertains to the maximum distance between the two components used to calculate the jet ridge line width (see text for details).}
\label{tab:jet_width_statistics_z_1}
\begin{tabular}{l| l l| l l| l l}
\multicolumn{7}{l}{}\\
\multicolumn{7}{l}{Jet Width ($^\circ$),  $\mathbf{0 < z}$}\\
\hline						
\textbf{Types}	&\multicolumn{2}{c}{\textbf{FSRQ}}&\multicolumn{2}{c}{\textbf{BL}} &\multicolumn{2}{c}{\textbf{RG}}\\
\hline			
\textbf{Distance (pc)}	&\textbf{All}	&$\mathbf{< 20}$ 		&\textbf{All}		&$\mathbf{< 20}$		&\textbf{All}		&$\mathbf{< 20}$	\\
\hline	\multicolumn{7}{l}{}                                                                        \\
\#	                &343		&251  	&63		    &51	    &95		    &72	    \\
\textbf{Average}	&12.0		&10.8 	&19.3		&18.2	&16.3		&16.5	\\
\textbf{Error}	    &0.4		&0.5  	&1.8		&1.8	&1.0		&1.2	\\
\textbf{Median}	    &8.7		&7.0  	&12.5		&12.5	&13.0		&13.2	\\
\textbf{Error}      &0.5        &0.6    &2.2        &2.2    &0.7        &0.8    \\
\textbf{Max	}       &86.4		&86.4 	&93.2		&51.9	&72.8		&72.8	\\
\textbf{Min}	    &0.1		&0.1  	&0.7		&0.9	&0.2		&0.2	\\
\multicolumn{7}{l}{}\\
\multicolumn{7}{l}{Jet Width ($^\circ$),  $\mathbf{0 < z < 1}$}\\
\hline						
\textbf{Types}	&	\multicolumn{2}{c}{\textbf{FSRQ}}	     	&\multicolumn{2}{c}{\textbf{BL}}		&\multicolumn{2}{c}{\textbf{RG}}\\
\hline		
\textbf{Distance (pc)}	&\textbf{All}	&$\mathbf{< 20}$ 		&\textbf{All}		&$\mathbf{< 20}$		&\textbf{All}		&$\mathbf{< 20}$	\\
\hline\multicolumn{7}{l}{}\\
\#		            &91	    &76   			&58		    &49		    &83		    &62	    \\
\textbf{Average}	&11.1	&10.5 			&20.4		&18.4		&16.6		&16.4	\\
\textbf{Error}		&0.8	&0.9  			&2.0		&1.9		&1.2		&1.4	\\
\textbf{Median}		&7.5	&6.5  			&13.8		&12.5		&12.3		&10.8	\\
\textbf{Error}      &1.0    &1.4            &2.5        &2.4        &1.5        &1.7    \\
\textbf{Max}		&47.7	&47.7 			&93.2		&51.9		&72.8		&72.8	\\
\textbf{Min}		&0.1	&0.1  			&0.9		&0.9		&0.2		&0.2	\\
\multicolumn{7}{l}{}\\
\hline
\end{tabular}
\end{center}
\end{table*}

\begin{figure}[h]
\begin{center}
  \includegraphics[width=0.5\textwidth,angle=0]{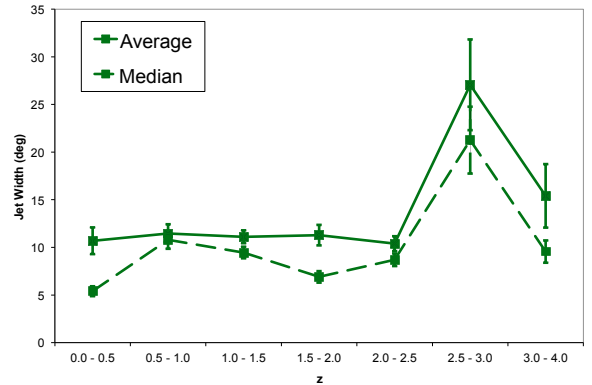}
  \caption{Jet width for CJF FSRQs, as a function of redshift. We use 0.5 redshift binning to calculate averages (continuous line) and median values (dashed line). Number of sources per bin: 38 (0-0.5), 53 (0.5-1), 106 (1-1.5), 61 (1.5-2), 55 (2-2.5),  18 (2.5-3), 12 (3-4).}
  \label{fig:jet_width_FSRQ_z}
\end{center}
\end{figure}

The jet ridge line width appears to be independent of redshift. To test this further, we use the sub-sample of CJF FSRQs to calculate average and median width values for different redshift bins (Fig. \ref{fig:jet_width_FSRQ_z}). Out to z=2.5 the jet ridge line width remains fairly stable, with the median value showing larger changes. For the last two bins, the number of sources contained drops drastically (as reflected by the much larger error bars) and therefore it is difficult to conclude whether the strong increase in both average and median values is a true or a spurious effect. It should be noted that this sharp increase coincides with the epoch of maximum nuclear and star-formation activity in the Universe (at $z\sim2-3$, e.g., \citealt{Hasinger1998}; \citealt{Madau1996}). Focusing on the redshift bin [0,1], from Table \ref{tab:jet_width_statistics_z_1} we see that BL Lac jet ridge lines appear significantly wider than their FSRQ counterparts ($20.4\pm2.0^\circ$, compared to $11.1\pm0.8^\circ$; $>4\sigma$). The same behavior is seen when looking at the median values (albeit at the $2\sigma$ level). Considering the values when the maximum distance limit between the $\theta_{max}$ and $\theta_{min}$ components  is assumed, BL Lacs remain substantially (at a $\sim4\sigma$ difference level) wider. Radio galaxies show similar jet widths to BL Lacs within the $1\sigma$ uncertainties.

\begin{figure*}[htb]
\begin{center}
  \includegraphics[width=0.45\textwidth,angle=0]{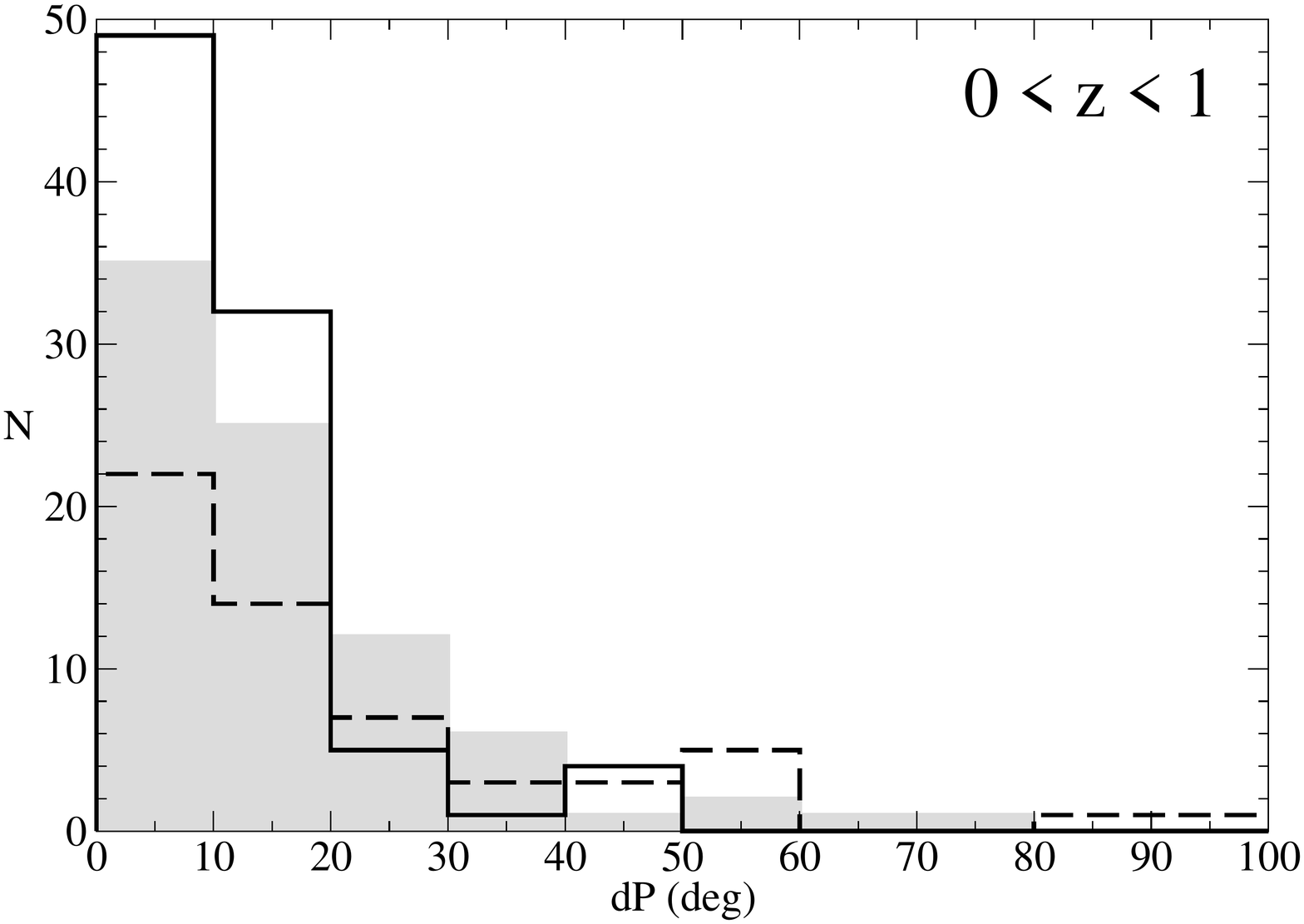}
   \includegraphics[width=0.445\textwidth,angle=0]{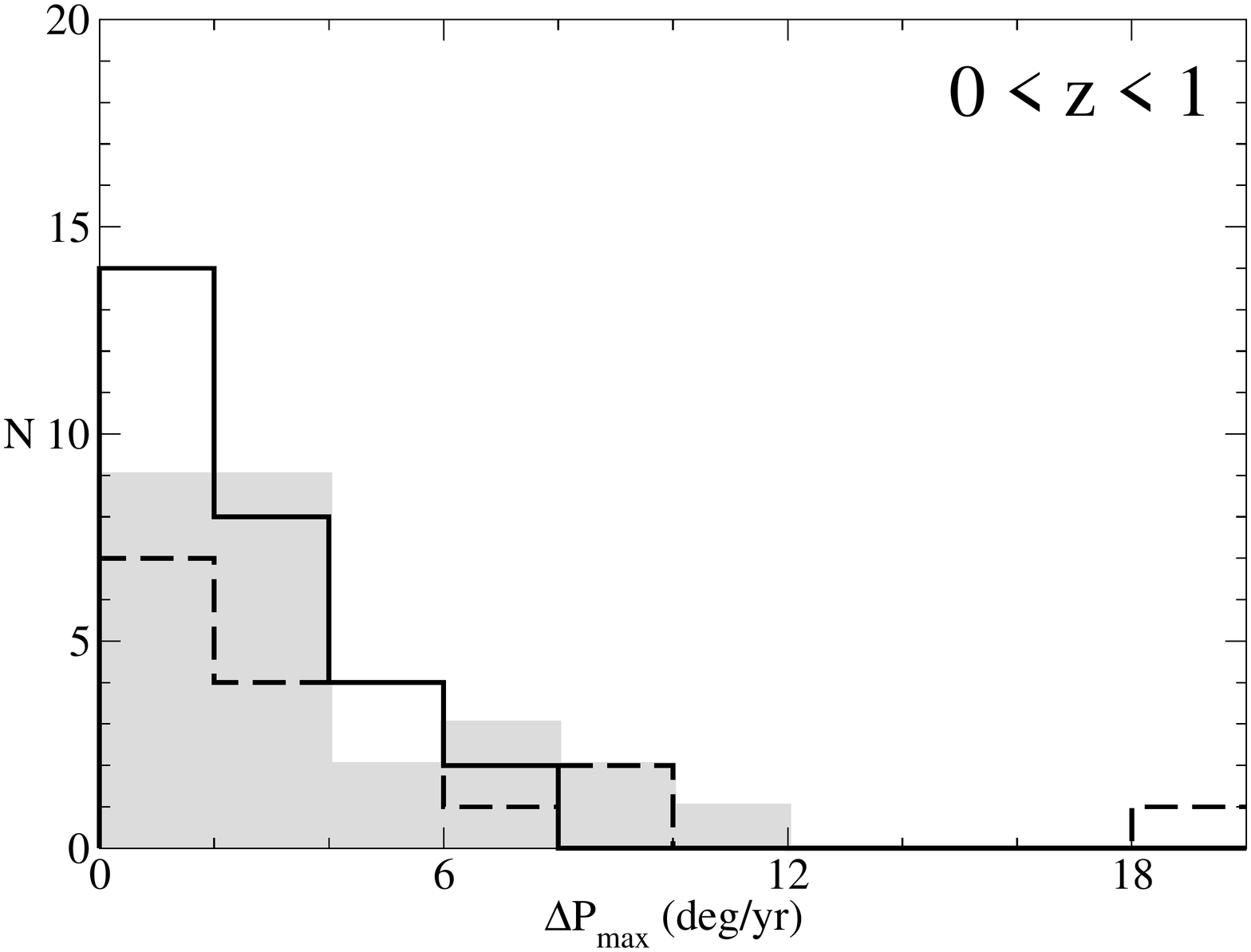}
  \caption{Jet ridge line width (left) and width evolution (right) distributions for BL Lacs (dashed line), FSRQs (solid line), and RGs (grey blocks). For this histogram sources with redshifts in the [0,1] bin are used. Sources flagged as ``NO'' are excluded.}
  \label{fig:jet_width_types_0_z_1}
\end{center}
\end{figure*}

In Fig. \ref{fig:jet_width_types_0_z_1} (left) we show the distribution of jet ridge line widths for BL Lacs, FSRQs, and RGs, in the redshift bin [0,1]. It can be seen that all three classes show similar distributions, with their maxima situated at around 10 degrees. The distribution of BL Lac appears to be wider and extending to larger widths, compared to FSRQs. FSRQs show a rather more contained distribution to lower width values than BL Lacs. For the 125 epochs where CJF sources show apparent jet widths $>20$ degrees, we find 14 BL Lacs, 27 FSRQs, and 13 RGs. When taking into account the total number of each type of object, we calculate that 47.3\% of BL Lacs have jets wider than 20 degrees, as opposed to only 13.6\% and 25\% for FSRQs and RGs, respectively.

We apply the two-sample Kolmogorov-Smirnoff (K-S) test to our data to see whether the three classes of AGN, BL Lacs, FSRQs, and RGs, are indeed different in their jet ridge line widths. Comparing the jet ridge line widths of BL Lacs and FSRQs (no redshift binning)the K-S test gives a $0.7\cdot10^{-3}$\% probability that these two sub-samples originate from the same parent distribution. Comparing FSRQs with RGs, we get a probability of $0.7$\%. The K-S test does not reject the null hypothesis (22.1\% probability) concerning the comparison between BL Lacs and RGs. We also compare the sub-samples of BL Lacs, FSRQs, and RGs in the redshift bin [0,1]. In this case the K-S test gives a probability of 1\% that BL Lacs and FSRQs are drawn from the same parent sample. For BL Lacs and RGs the test again does not reject the null hypothesis. Finally, we also apply the K-S test after we assume the 20 pc distance constraint between the $\theta_{max}$/$\theta_{min}$ components.  In this case we get a 4.4\% probability for BL Lacs and FSRQs, while for BL Lacs and RGs the test once more does not reject the null hypothesis. The latter comparison is affected by low-number statistics. We can conclude that BL Lacs show a significantly different distribution of jet ridge line widths compared to FSRQs. In both average and median values, BL Lacs show substantially wider jets than both FSRQs and RGs.

\subsection{Apparent jet width evolution, $\Delta P$, statistics}
\label{sec:rl.width_ev}
As mentioned before, component identification is unimportant when working with jet ridge lines. While this is true, the actual detection of the same number of components across different epochs does bear some importance for our results. The effect of a vanishing component (either because of diminishing flux, bad observing conditions, or otherwise flawed data) will induce a spurious width change, affecting the overall statistics of the sample. Although in the case of diminishing flux, one could argue that this change is indeed intrinsic and therefore important, usually the latter two effects dominate.  Moreover, as we already mentioned in Sect. \ref{sec:data}, the given sensitivity and dynamic range of the observations define our results. For much more sensitive observations, jets would probably appear rather wider and with a more complex structure than what we find here.

To take this effect into account we flag our data accordingly. The best quality data (``OK") are those for which a width change is calculated between epochs with the same number of components identified per source. As moderate quality data (``!") are flagged those for which the width change is calculated between epochs with $\pm1$ number of components identified per source. If the number of components varies by 2 or more, the data are flagged as bad (``NO"). In this way we can still take into account the ejection of new components. Within the time spans between most epochs ($\sim2$ years) the appearance of more than one new component is not expected. In Table \ref{tab:flag_width_ev} the number of epochs for each class of object with each of the corresponding flags is shown. As can be seen, FSRQs and RGs have the best quality data, with only a 24\% and 27\% respectively of the data flagged as not ``OK" (``!" and ``NO" flag). In contrast BL Lacs have almost half their data (47\%) flagged accordingly. This reflects the difficulty of a consistent component model-fitting of BL Lac jets across epochs. This is a result of (1) the variable nature of these objects, (2) the intrinsically fainter nature of BL Lacs, and (3) the possible shorter length of their jets that leads to confusion effects between components (compared for example to FSRQ jets).

\begin{table}
\begin{center}
\caption{Three different types of flagging for our data, along with their definition. For each category the corresponding number for each class of object is shown. For an epoch i, each source has $N_{i}$ number of components identified.}
\label{tab:flag_width_ev}
\begin{tabular}{l| c| l l l}
\multicolumn{5}{l}{}\\
\hline						
	&\textbf{Definition}		&\multicolumn{3}{c}{\textbf{\# of epochs}}	\\
	&			&\textbf{FSRQ}	&\textbf{BL}	&\textbf{RG}     \\
\hline\multicolumn{5}{l}{}\\
\textbf{OK}	&$N_{l}=N_{l+1}$		    &173  &26	&54     \\
\textbf{!}	&$N_{l}=N_{l+1}\pm1$	    &53	  &21	&15     \\
\textbf{NO}	&$N_{l}\geq N_{l+1}\pm2$	&2	  &2	&5      \\
\multicolumn{5}{l}{}\\
\hline
\end{tabular}
\end{center}
\end{table}

We calculate a non-zero maximum apparent width evolution for 180 sources (sources flagged as ``NO" are not counted here). Of there, 51 show an evolution of their width larger than 4 deg/yr, while 17 have a $\Delta P_{max}>8$. We thus find that 28.3\% of the sources show significant evolution of their apparent jet width.

We compare the maximum values of the jet ridge line width evolution per source between BL Lacs, quasars, and radio galaxies, using no maximum distance limit between the $\theta_{max}$ and $\theta_{min}$ components for the width calculation and excluding data flagged as ``NO". The differences in average and median values for the different classes are not found to be statistically significant ($<3\sigma$). BL Lacs do however show a trend of having stronger width evolution than the other two classes. When checking the maximum and minimum values of each sub-sample, it becomes obvious that FSRQs and BL Lacs show much more extended distributions compared to RGs.

\begin{table*}[hbt]
\begin{center}
\caption{Statistical properties of the maximum jet width evolution (left) and jet linear evolution (right) distributions for FSRQs, BL Lacs, and RGs. Average, median , maximum, and minimum values are calculated for both all sources, as well as for sources in the redshift bin [0,1]. Sources flagged as ``NO" are excluded. Only the inner part of the jet is considered ($<40$ pc).}
\label{tab:jet_width_ev_max_z}
\begin{tabular}{l| l l l| l l l}
\multicolumn{7}{l}{}\\
\multicolumn{1}{l}{}&\multicolumn{6}{l}{\large{\textbf{Jet width evolution (max)} ($deg/yr$)}}\\
\hline
	&\multicolumn{3}{c|}{All}	&\multicolumn{3}{c}{$0 < z < 1$}\\			
\textbf{Types}	&\textbf{FSRQ}	&\textbf{BL}	&\textbf{RG}		&\textbf{FSRQ}	&\textbf{BL}	&\textbf{RG} \\
\hline\multicolumn{7}{l}{}\\
\#	            &114  		&26   		&35   			&29           	&19         &26          \\
\textbf{Average}&3.44 		&3.9 		&3.3 			&3.5  		    &4.1  		&3.7        \\
\textbf{Error}	&0.23 		&0.6 		&0.3 			&0.4  		    &0.7		&0.4        \\
\textbf{Median}	&2.26 		&2.3 		&2.5 			&2.42        	&2.8      	&2.73        \\
\textbf{Error}  &0.29       &0.7        &0.4            &0.5            &0.9        &0.5        \\
\textbf{Max}	&27.11		&18.32		&11.84			&24.39       	&18.32     	&11.84       \\
\textbf{Min}	&0.16 		&0.34 		&0.14 			&0.41        	&0.12      	&0.51        \\
\multicolumn{7}{l}{}\\
\hline
\end{tabular}\hspace{30pt}
\begin{tabular}{ l l l| l l l}
\multicolumn{6}{l}{}\\
\multicolumn{6}{l}{\large{\textbf{Jet linear evolution} ($pc/yr/comp$)}}\\
\hline
	\multicolumn{3}{c|}{$0 < z$}	&\multicolumn{3}{c}{$0 < z < 1$}\\			
\textbf{FSRQ}	&\textbf{BL}	&\textbf{RG}		&\textbf{FSRQ}	&\textbf{BL}	&\textbf{RG} \\
\hline\multicolumn{6}{l}{}\\
171  			&25   		&41   			&44           		    &21         	&35         \\
0.470		&0.38		&0.34			&0.51 		    &0.38 		&0.32      \\
0.016		&0.05		&0.03			&0.04 		    &0.05		&0.03      \\
0.414		&0.24		&0.28			&0.50       	  	    &0.24     	&0.25      \\
0.020      		&0.06       		&0.04           		&0.04           	    &0.06       	&0.04       \\
1.910 		&1.081		&0.989			&1.910         	    &1.081     	&0.989      \\
0.028		&0.038		&0.028			&0.028       	    &0.038     	&0.028      \\
\multicolumn{6}{l}{}\\
\hline
\end{tabular}
\end{center}
\end{table*}

In Fig. \ref{fig:jet_width_types_0_z_1} (right) we show the distributions for the maximum jet ridge line width evolution for FSRQs, BL Lacs, and RGs. We use a redshift bin of [0,1] and exclude all data flagged as ``NO" (this corresponds to the last three columns of the left part of Table \ref{tab:jet_width_ev_max_z}). BL Lac objects show a wide distribution of jet width evolution values, with a primary maximum at 3 degrees/yr and extending out to 18 deg/yr. Radio galaxies show a somewhat similar behavior with a plateau between 2 and 4 degrees/yr. Quasars, in contrast, show a confined distribution with a strong maximum around 3 degrees/yr and all values contained between 0 and 10 degrees/yr. We apply the K-S test to see whether these distribution are significantly different. Concentrating on the comparison in the [0,1] redshift bin, the K-S test does not reject the null hypothesis (33.8\% probability). Calculating the relative occurrence of the different classes, we find that 12.5\% of BL Lacs show $\Delta P_{max}>8$, as opposed to 4\% and 5.7\% for FSRQs and RGs, respectively.

\subsection{Apparent jet linear evolution, $\Delta\ell$, statistics}
Focusing on the final measure of the jet ridge line kinematics, as studied in this paper, we want to investigate whether the stationarity of components, as observed in the case of S5 1803+784, is commonplace among other BL Lacs. We use the linear evolution measure, as described in Sect. \ref{sec:data}, to do this. We follow the same procedure as previously, to check the statistics of the individual classes.

In Table \ref{tab:jet_width_ev_max_z} (right) we give the statistical properties of the total linear evolution of the jet ridge line (measured in parsecs per unit time and component) distributions for FSRQs, BL Lacs, and RGs. On average, both BL Lacs and RGs show weaker evolution of their jet ridge lines compared to FSRQs ($\sim2\sigma$ difference). Looking at the median values, BL Lacs and RGs show the least evolution compared to FSRQs ($0.24\pm0.06$, $0.28\pm0.04$, and $0.414\pm0.020$ pc/yr/comp respectively; $\sim3\sigma$ difference). We also study the statistics of the sub-sample of CJF sources in the [0,1] redshift bin (also in Table \ref{tab:jet_width_ev_max_z}). The behavior remains the same as before, with differences between BL Lacs and FSRQs becoming more pronounced (with a $\sim4\sigma$ difference in median values).

\begin{figure*}[htp]
\begin{center}
  \includegraphics[width=0.5\textwidth,angle=0]{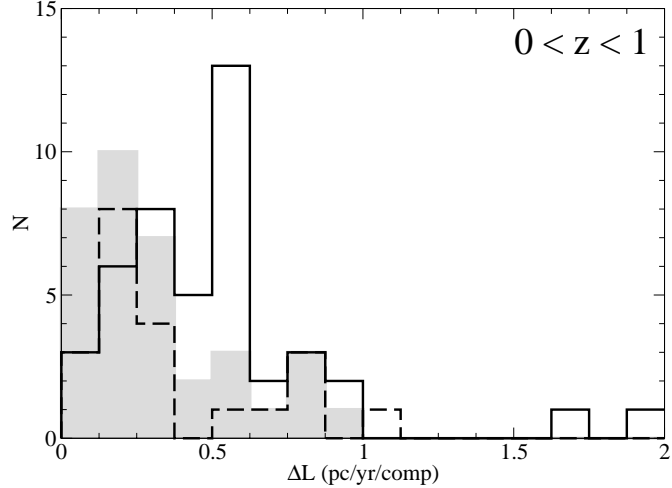}
  \caption{Distribution of the total linear evolution of the CJF jet ridge lines for FSRQs (solid line), BL Lacs (dashed line), and RGs (grey blocks). We use sources with redshift lower than 1.}
  \label{fig:jet_linear_ev_histo}
\end{center}
\end{figure*}

We now turn to the actual distribution of the total jet linear evolution for the three types of objects, shown in Fig. \ref{fig:jet_linear_ev_histo}. We once again only take into account sources in the redshift bin [0,1]. Quasars show a pronounced maximum around 0.63 pc/yr/comp, extending out to 2 pc/yr/comp. In contrast, both BL Lacs and RGs show their distribution maxima around 0.25 pc/yr/comp. Quasars appear to have a wider distribution of total linear evolution values, also showing the highest maximum values among all three types (see Table \ref{tab:jet_width_ev_max_z}). Interestingly, RGs and BL Lacs, in total, show very similar distributions. Once again we employ the K-S test to compare the distributions. For the whole sample (independent of redshift constraints), the K-S test gives a probability of 4.3\% that BL Lacs and FSRQs are drawn from the same parent population. For BL Lacs and RGs the test does not reject the null hypothesis. Focusing on the $0<z<1$ sub-sample, we get a marginal 5.4\% probability that BL Lacs and FSRQs stem from the same parent sample. These results give a positive answer as to whether BL Lacs show less absolute linear evolution of their jets compared to their FSRQ counterparts. A somewhat surprising result concerns the comparison between BL Lacs and RGs.

\subsection{Jet ridge line properties and variability}

\citet{Karouzos2010} gathered variability information from the literature for the CJF sources, focusing on the long timescale ($\sim1-10^{1}$ years) variability, often linked to a helical jet structure and/or a binary AGN core. We use this information to define a subsample of 40 CJF variable sources and study their properties in terms of the measures discussed in Sect. \ref{sec:data}. Given the small number of sources/epochs for this investigation, we impose no additional constraints to our sub-sample (e.g., redshift bins).

\begin{figure*}[htp]
\begin{center}
\begin{tabular}{c c}
  \includegraphics[width=0.45\textwidth,angle=0]{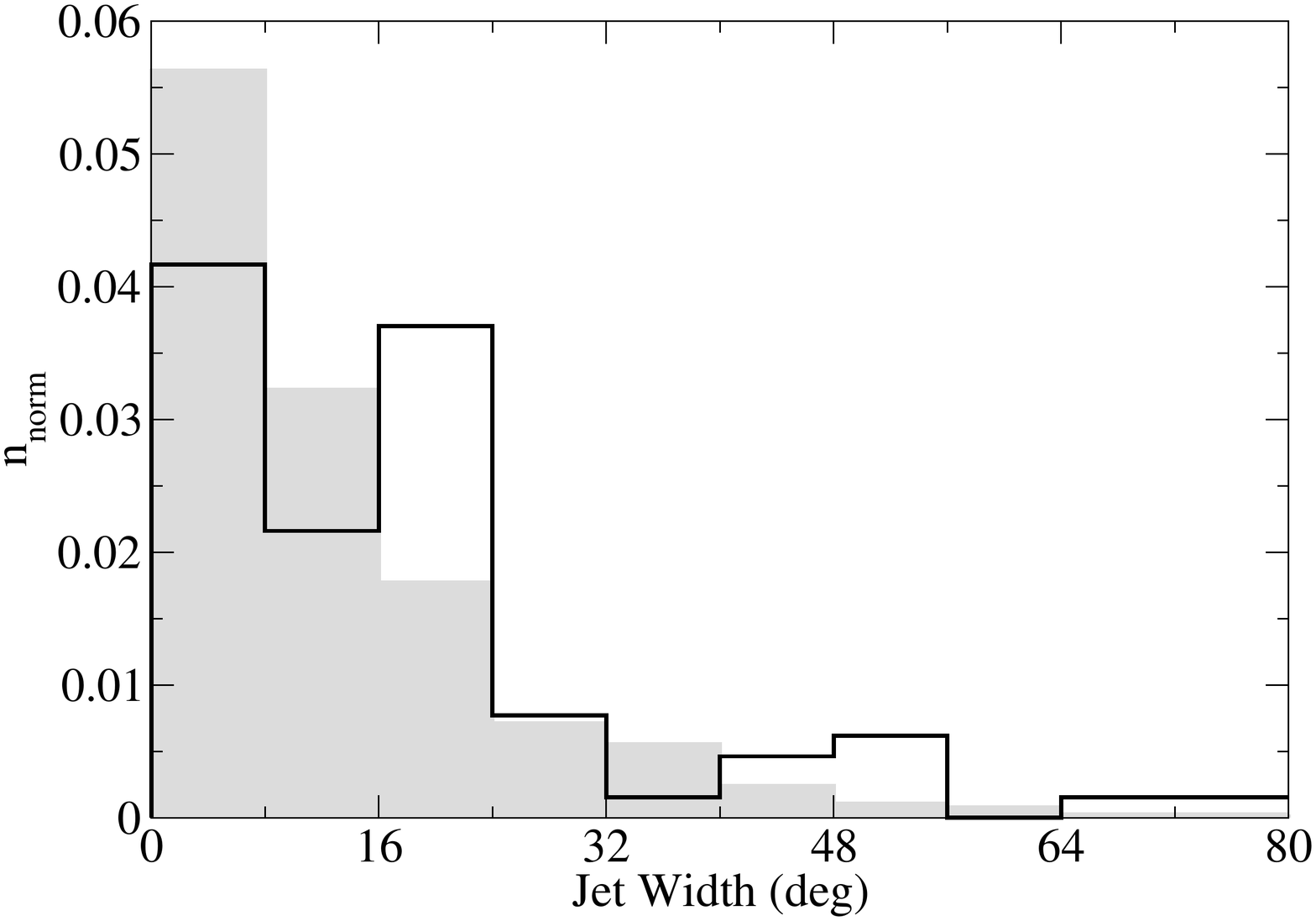}
   &\includegraphics[width=0.39\textwidth,angle=0]{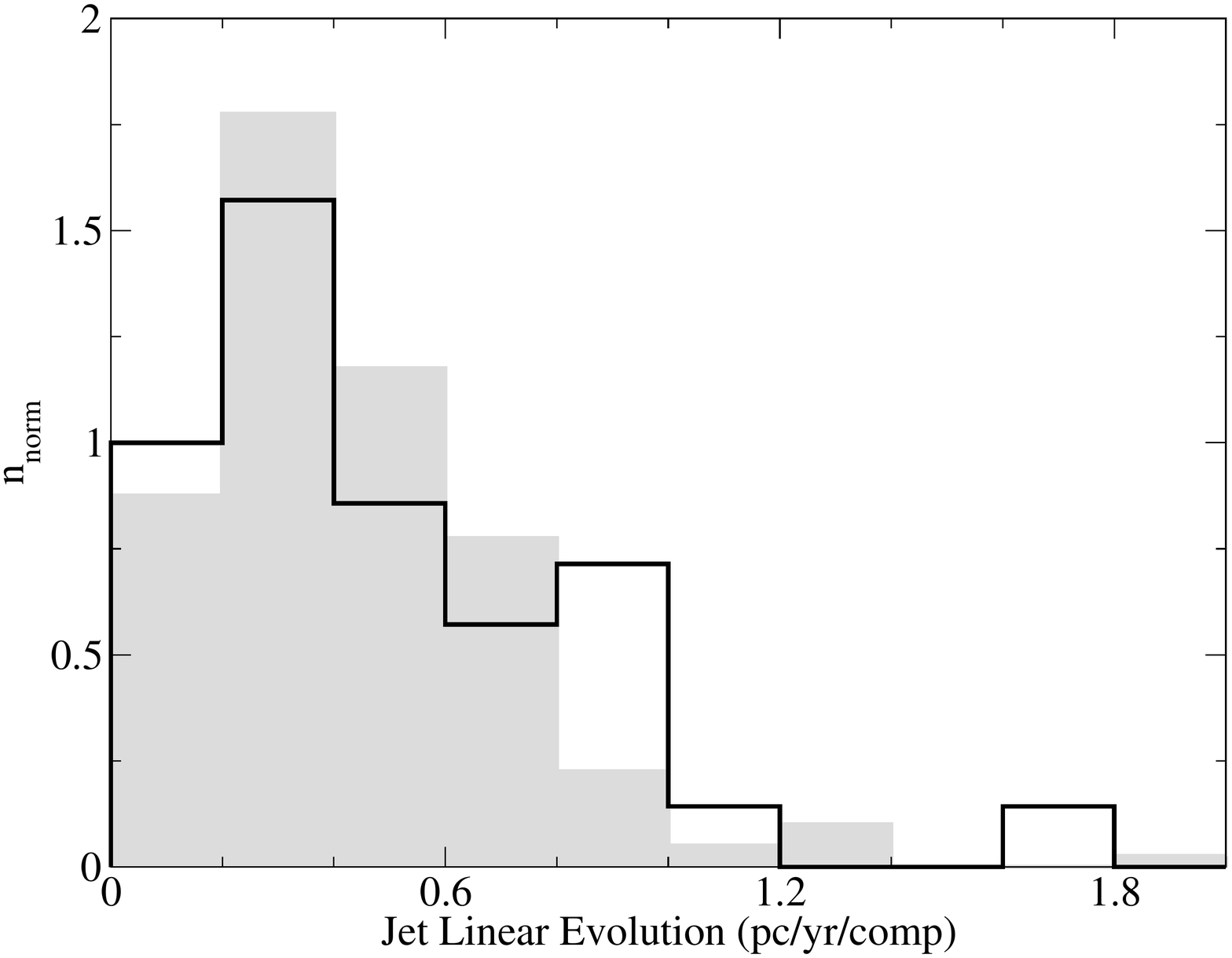}
   \end{tabular}   
  \caption{Jet ridge line width (left) and linear evolution (right) distributions for the sub-sample of variable sources from \citet{Karouzos2010} (black line) and for the rest (non-variable) CJF sample (grey blocks). Due to the big difference in absolute numbers for the two samples, we have normalized to unity surface area. No redshift or distance constraints are assumed.}
  \label{fig:jet_width_histo_var}
\end{center}
\end{figure*}

The sub-sample of variable sources exhibits both higher average ($18.3\pm1.3^\circ$, compared to $14.4\pm0.4^\circ$; $\sim3\sigma$ difference) and median ($15.7\pm0.9^\circ$, compared to $9.70\pm0.26^\circ$; $6\sigma$ difference) width values. In Fig. \ref{fig:jet_width_histo_var} (left) we plot the jet ridge line width distribution for the sub-sample of variable sources, as well as for the rest of the CJF. We see that the parent (non-variable\footnote{We underline the fact that the number of variable sources is explicitly regulated by the availability, or lack thereof, of the appropriate observations (long-timescale variability can be uncovered only by means of extensive monitoring of a source). Some of the sources labeled here as ``non-variable" are probably variable but not observed as such.}) CJF sample shows strong maximum in the lowest bin $[0^\circ,8^\circ]$ with a decreasing trend. In contrast, the sub-sample of variable sources show two maxima in bins $[0^\circ,8^\circ]$ (main) and $[16^\circ,24^\circ]$ (secondary), with a secondary to primary ratio of $\sim0.9$. Variable CJF sources indeed show a more extended distribution of jet ridge line widths, shifted towards higher values compared to the parent CJF sample. A K-S test gives a very low (0.3\%) probability that the two sub-samples are drawn from the same parent population.

\begin{table*}
\begin{center}
\caption{Statistical properties of the jet width, maximum jet width evolution, and jet linear evolution for the sub-sample of variable CJF sources (see text for definition of variability). Average and median are calculated for all sources, independent of redshift. Sources flagged as ``NO" are excluded. Only the inner part of the jet is considered ($<40$ pc). For each measure the upper row gives average values, while the median values are found in the lower rows.}
\label{tab:width_ev_lin_var}
\begin{tabular}{c | c | c | c| c }
\multicolumn{1}{c|}{} &\textbf{All}	&\textbf{FSRQ}	&\textbf{BL Lac}	&\textbf{RG}	\\
\hline
\textbf{P}			 &18.3$\pm$1.3	   &22.0$\pm$2.7	&17.1$\pm$1.7		&12.9$\pm$2.1\\
(deg)			 &15.7$\pm$1.6	   &17.8$\pm$3.3	&13.5$\pm$2.1		&13.4$\pm$2.6\\
\hline
\textbf{$\Delta$P}	&5.5$\pm$0.9	   &8.4$\pm$2.3	&4.5$\pm$0.9		&1.9$\pm$0.6\\
(deg/yr)			&3.1$\pm$1.1	   &5.0$\pm$2.9	&2.5$\pm$1.1		&1.6$\pm$0.7\\
\hline							
\textbf{$\Delta\ell$}	&0.49$\pm$0.05 &0.61$\pm$0.06	&0.36$\pm$0.07	&0.35$\pm$0.17\\
(pc/yr/comp)		&0.38$\pm$0.06 &0.53$\pm$0.08	&0.23$\pm$0.08	&0.14$\pm$0.21\\
\end{tabular}
\end{center}
\end{table*}

BL Lacs appear to maintain the same average and median width as their parent sample (within errors). FSRQs however appear to have both much higher average ($22.0\pm2.7^\circ$, compared to $10.8\pm0.5^\circ$; $4\sigma$ difference) and median values than their parent sample ($17.8\pm1.6^\circ$, compared to $7.00\pm0.28^\circ$; $4\sigma$ difference). Variable RGs showmarginal evidence for lower average ($12.9\pm2.1^\circ$, compared to $16.5\pm1.2^\circ$) but similar median value.

For $\Delta P$, compared to the parent sample, the sub-sample of variable sources exhibits both higher average ($5.5\pm0.9^\circ/yr$, compared to $3.51\pm0.18^\circ/yr$) and median values ($3.1\pm0.4^\circ/yr$, compared to $2.37\pm0.11^\circ/yr$), albeit at a low ($2\sigma$) significance level. Variable BL Lacs appear to maintain both the same average width as their parent sample (within errors), as well as the same median value ($2.5\pm0.5^\circ/yr$, compared to $2.3\pm0.3^\circ/yr$). FSRQs appear to have both higher average ($8.4\pm2.3^\circ/yr$, compared to $3.44\pm0.22^\circ/yr$; $2\sigma$ difference) and median values ($5.0\pm1.0^\circ/yr$, compared to $2.26\pm0.13^\circ/yr$; $2.5sigma$) than their parent sample. Variable RGs show lower average ($1.9\pm0.6^\circ/yr$, compared to $3.3\pm0.3^\circ/yr$) but similar median values ($1.6\pm1.0^\circ/yr$, compared to $2.48\pm0.20^\circ/yr$).

For $\Delta\ell$, compared to the parent sample, the sub-sample of variable sources does not exhibit significantly different average and median values. For both samples the total maximum of the distribution occurs in the [0.2,0.4] bin, with the variable CJF sources however showing a less strong maximum compared to the non-variable sample (see Fig. \ref{fig:jet_width_histo_var}, right). In contrast to the non-variable sources, the variable sample shows a secondary maximum in the bin [0.8,1] pc/yr/comp implying a broader distribution of $\Delta\ell$ for variable sources. The K-S test does not reject the null hypothesis (23.8\% probability). BL Lac objects appear to maintain the same average and median linear evolution as their parent sample (within errors). Variable quasars appear to have both higher average ($0.615\pm0.063$, compared to $0.470\pm0.016$; $>4\sigma$ difference) and median values ($0.53\pm0.04$ pc/yr/comp, compared to $0.414\pm0.020$ pc/yr/comp) than their parent sample. Variable RGs show similar average values (within errors) but lower median value ($0.135\pm0.003$ pc/yr/comp, compared to $0.28\pm0.04$ pc/yr/comp).

There seems to be a strong link between the kinematic properties studied here and the variability observed in these sources. For all three measures BL Lacs appear to maintain their statistical behavior between variable and parent sample, whereas variable quasars tend to lean towards a ``BL Lac-like" behavior. It should be noted however that, especially for RGs, this comparison is problematic due to the small number of sources (3 RGs).

\subsection{Jet kinematics and source luminosity}
In addition to the above, we also look for a possible correlation between the kinematic and morphological measures described in Sec. \ref{sec:data} and the radio luminosity of the sources, a value probably less heavily dependent on the jet structure/morphology.

\begin{figure*}[htp]
\begin{center}
  \includegraphics[width=0.45\textwidth,angle=0]{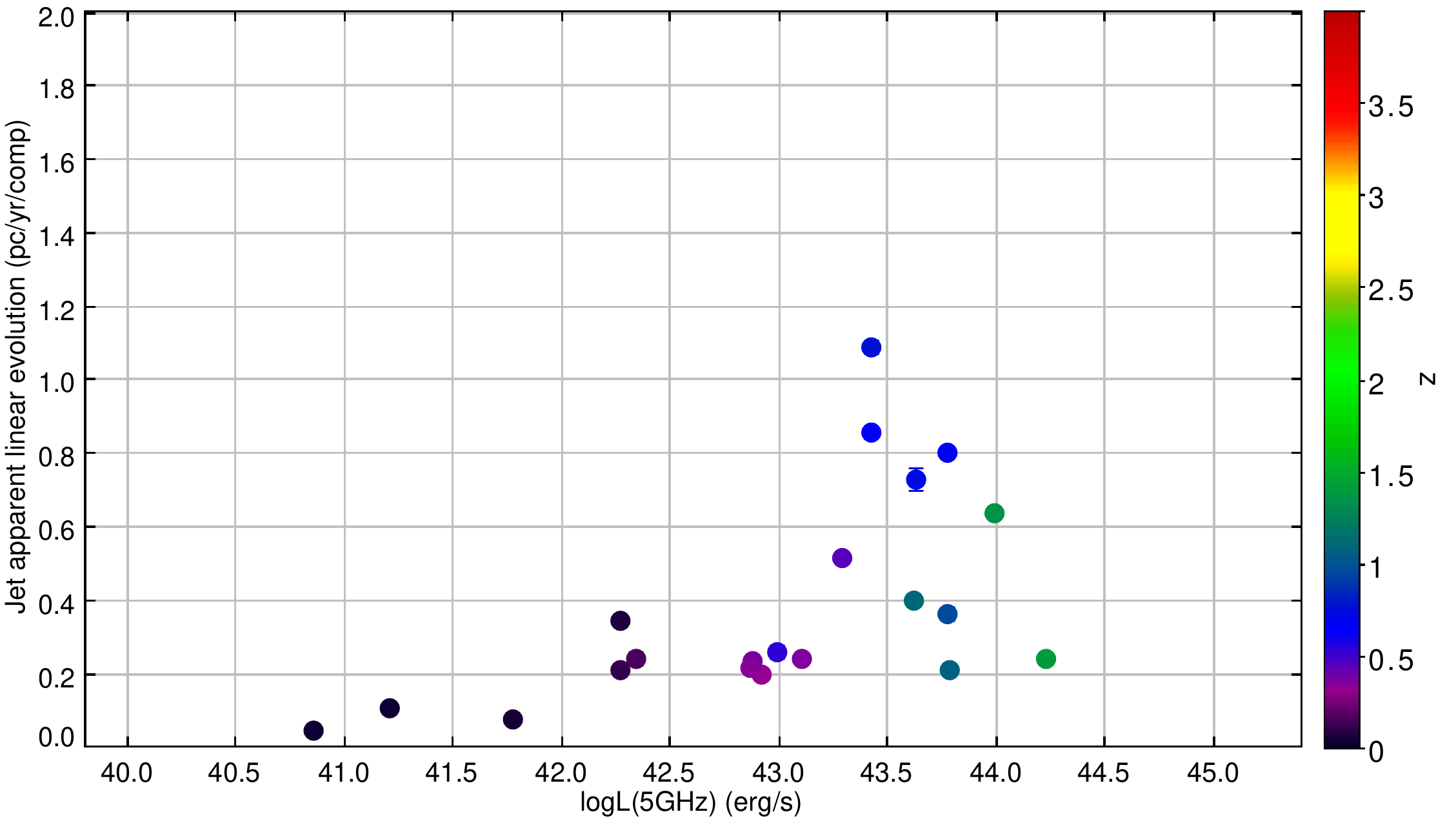}
  \includegraphics[width=0.45\textwidth,angle=0]{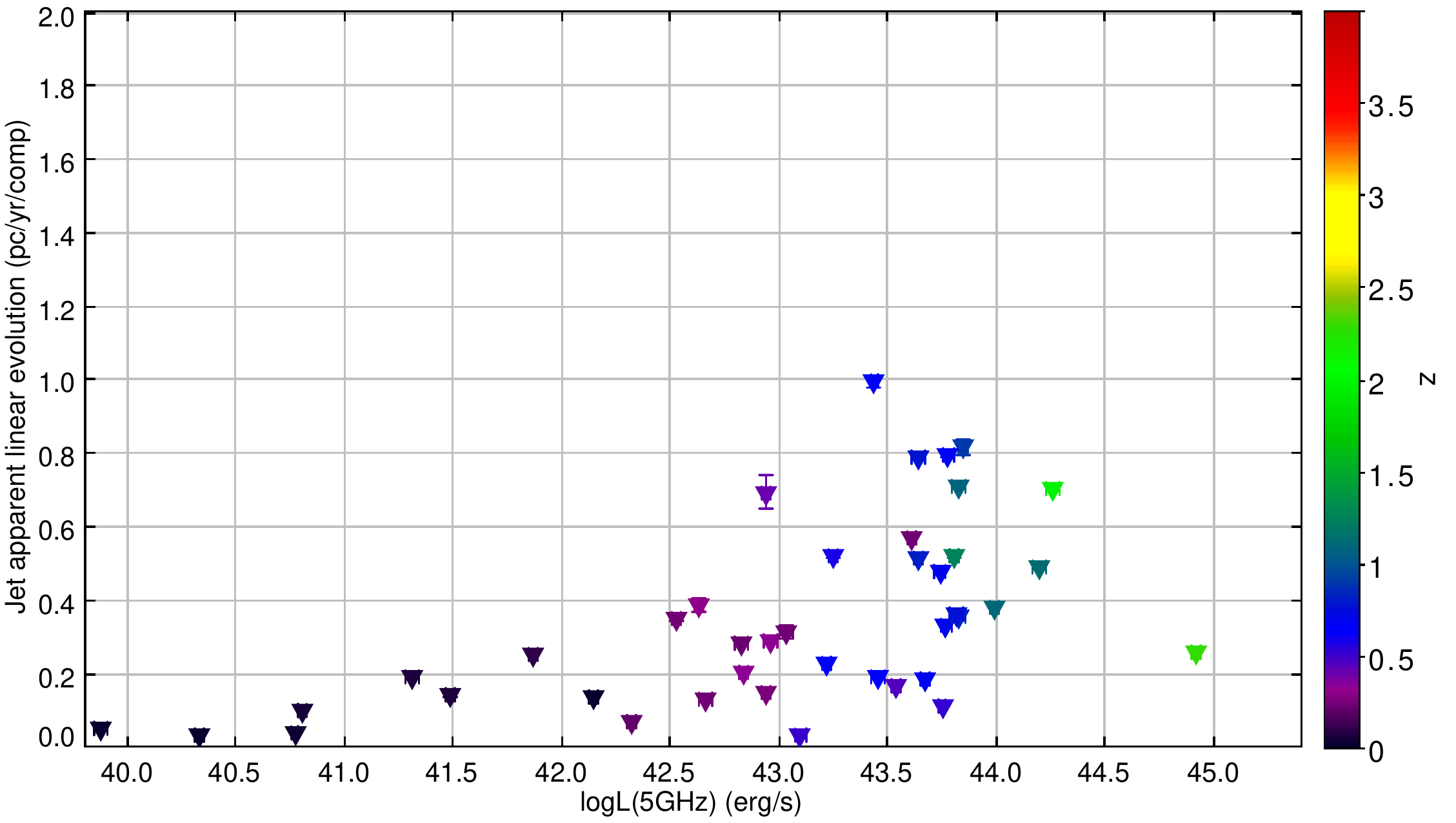}\\
  \includegraphics[width=0.5\textwidth,angle=0]{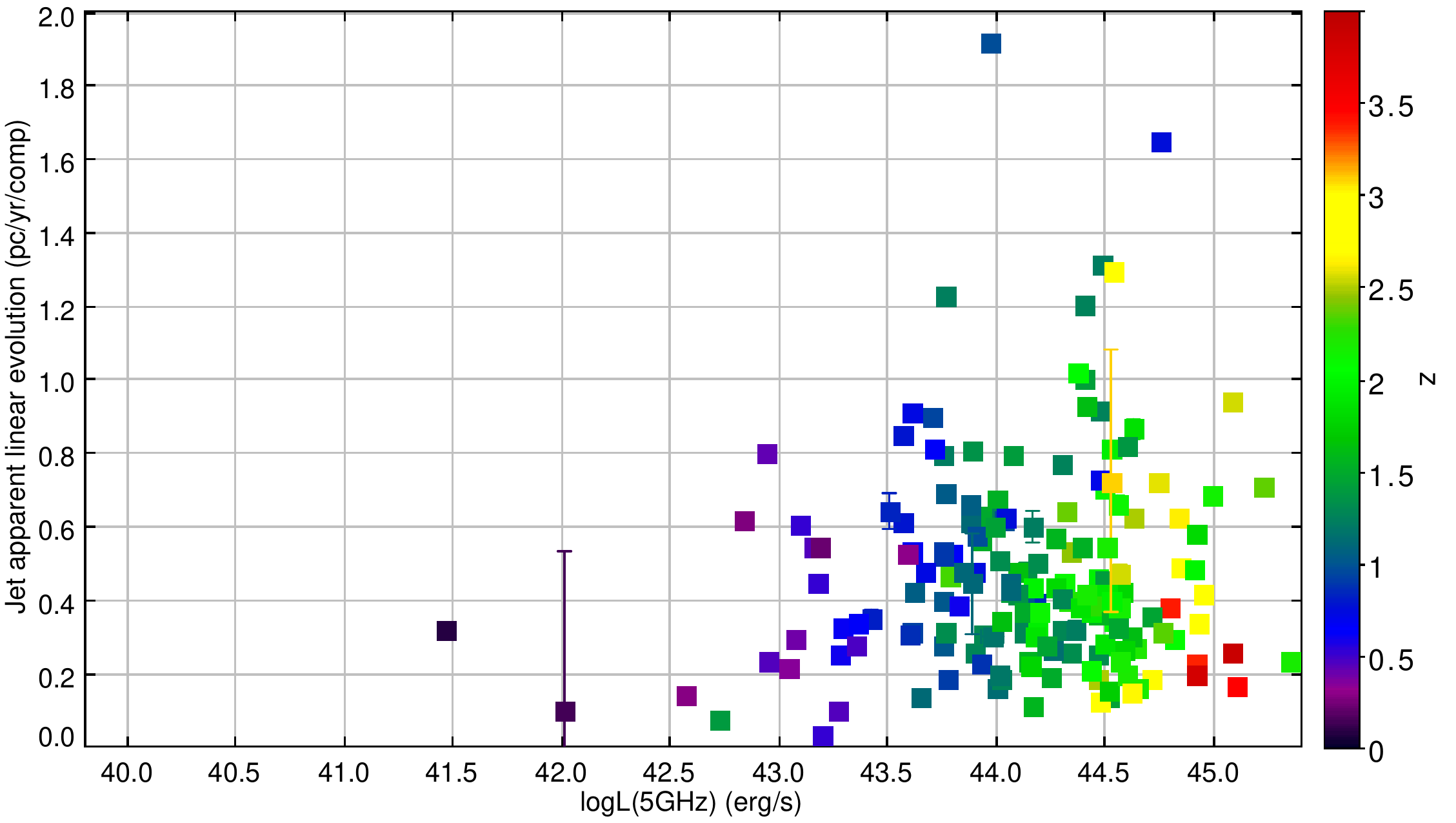}
  \caption{Total linear evolution of the jet ridge lines for the BL Lac and RG sub-samples  (up; left and right, respectively) and the FSRQ sub-sample (down) as a function of the total source luminosity at 5 GHz. Quasars are shown with squares, BL Lacs with circles, and RGs with triangles. We use a color gradient to denote different redshifts.}
  \label{fig:jet_linear_ev_lum}
\end{center}
\end{figure*}

In Fig. \ref{fig:jet_linear_ev_lum} we show the total linear evolution of the CJF jet ridge lines as a function of the core luminosity of the source at 5 GHz (from VLA observations). It can be immediately seen that a correlation exists between radio luminosity and total linear evolution of the jet ridge line. This is seen as an upper envelope, above which no sources are found. More strongly evolving jets inhabit brighter sources. Conversely, low luminosity sources appear to show the smallest values of linear displacement of their jets. A similar correlation has been reported between maximum apparent jet-component speed and radio luminosity (\citealt{Cohen2007}; \citealt{Britzen2008}; \citealt{Lister2009c}). \citet{Britzen2008} discuss this correlation between apparent speeds and luminosity in terms of a redshift induced effect. As can be seen in Fig. \ref{fig:jet_linear_ev_lum}, there is an indication that this is not an effect caused by redshift (e.g., CJF sources at the highest redshifts ($>3$), albeit only being a few, all appear to have rather weak linear evolution of their jet ridge lines). Further arguments against a redshift induced effect dominating this correlation can be found in \citet{Britzen2008}. It should be noted that small-number statistics might be affecting our results. Both for low luminosity sources, as well as high redshift sources, we have a small number of objects available. That hinders the global validity of what is seen in Fig. \ref{fig:jet_linear_ev_lum}.

Concerning differences between FSRQs, BL Lacs, and RGs, no strong effects can be seen in this context. The low-luminosity tail of the distribution is predominantly populated by RGs and BL Lacs, with FSRQs mainly dominating the higher luminosities. \citet{Cohen2007} use Monte Carlo simulations to show that such an envelope can be explained in terms of ``aspect curves", essentially supporting the beaming model of relativistic jets as an explanation for this effect. In Fig. \ref{fig:jet_linear_ev_lum} (right) we plot the same correlation but only for the RGs in our sample. They form the same upper envelope, with more strongly evolving jets appearing in more luminous RGs. In the context of the unification scheme, where RGs are believed to be observed at large angles to their jet axes, Fig. \ref{fig:jet_linear_ev_lum} indicates that the correlation between total linear evolution of the jet ridge line and the luminosity can not be attributed only to geometric effects.

\section{Discussion}
\label{sec:discussion}

Given the flux-limited nature of the CJF sample, we are interested in investigating possible factors influencing the results presented in this paper. These can be briefly summarized in the following:
\begin{itemize}
\item relative number of FSRQs and BL Lacs
\item CJF redshift range
\item jet component identification
\item projection effects
\end{itemize}
We shall address each of these points separately.

\subsubsection{Total and absolute source numbers}
Compared to other samples of similar nature (e.g., MOJAVE), the CJF contains a larger number of sources, enabling the study of individual populations or sub-groups of sources (i.e., FSRQs, BL Lacs, etc.). It should be noted that, given the selection criteria of the sample and of course the relative cosmic abundance of these sources, quasars dominate the sample, with only a few BL Lacs and radio galaxies in comparison. This introduces a certain degree of small number statistics uncertainties for some of the comparisons in this paper. Future radio surveys with next-generation instruments will surely allow the definition of much larger samples that will address this problem.

\subsubsection{Redshift distribution}
The large redshift span of the CJF, as well as the uneven sampling of that range with respect to the three types of AGN influence our results. BL Lac objects are over-represented in the redshift range 0 to 2. Radio galaxies follow a similar behavior, with most sources in the redshift range [0,1], but a few sources at higher redshifts. Contrary to that, FSRQs show are over-represented in the redshift bin [1,2] and extending to the highest redshift values of the sample.

As we briefly discussed before, the fixed sensitivity and resolution of the arrays used allows us to have a direct comparison between the different CJF sources. On the same time however, it also essentially leads us to observe different scales of the jet for objects at different redshifts. It is expected that the properties, kinematics, and morphology of a jet are strong functions of the separation from the core. We therefore adopt redshift bins for all of the comparisons we do. Given the distribution of redshifts for the CJF the only bin we can use without losing most of the BL Lacs is the [0,1] one\footnote{It should also be noted that given the featureless spectrum, for which BL Lacs are famous, leads to more than 34\% of the CJF BL Lacs to not have available redshift information. When using redshift bins these objects are obviously excluded.}.

\subsubsection{Component identification}
The way that we study the width and width evolution of the jet is completely independent of component identification. The linear evolution of the jet ridge line on the other hand is an exception to this. For this measure we are forced to follow the given identification of components (from \citealt{Britzen2007a}) as we need some reference point to calculate absolute displacements. Despite the inevitable coupling of prior component identification to our $\Delta\ell$ measure, the fact that we include in this measure all components and all epochs ensures a treatment of the jet as a whole, evening out peculiarities or possible misidentifications of individual components.

Our measures are however sensitive to the number of components identified at a certain epoch. This is especially relevant for BL Lacs, given the nature of their jets. As we discussed shortly previously, BL Lacs are extremely variable sources. In parts this owns to the presumably small angle to our line of sight, resulting to beaming effects that change the flux of both the core and the jet. Secondly, we have shown that BL Lacs show indications of shorter jets than both FSRQs and RGs. This translates to higher difficulty in consistently model-fitting the jets of BL Lacs, as blending effects become more prominent as the viewing angle decreases. Moreover, given the strong variability of both core and jet, components might simply vanish as a result of diminishing luminosity or insufficient dynamic range of the observations. A further way to demonstrate this is by using the quality classification for jet components used by \citet{Britzen2008}. Jet component proper motions are classified as Q1 for the best quality data, and Q2 and Q3 for diminishing quality. The ratio $N(Q1)/(N(Q2)+N(Q3))$, where N(x) is the number of x quality components of that type of source, is 0.63 for FSRQs, 0.80 for RGs, and 0.41 for BL Lacs. Returning to our original point, we see therefore that the number of components across epochs in BL Lacs jets is more variable than both FSRQs and RGs. This introduces some uncertainty to our results.

\subsubsection{Projection effects}
The fact that the values calculated here are all projected onto the plane of the sky introduces some uncertainty and consequently a scatter in this statistical investigation. Although for this analysis and for the following discussion we have adopted the general AGN unification scheme, where BL Lacs and FSRQs are at the smallest viewing angles and radio galaxies are at larger viewing angles, the dispersion of the actual angles within each class introduces the aforementioned scatter in our statistics. 

An alternative path would be to try and deproject the jet values calculated here. Although certainly possible for a number of the CJF sources where a Doppler factor has been calculated (e.g., \citealt{Britzen2007b}, it would introduce further degrees of uncertainty to our analysis that are not so easily constrained. As we are mainly interested in the comparison between FSRQs and BL Lacs rather than the calculation of intrinsic jet properties, the deprojection is not essential for the results of this paper.

\subsection{Possible explanations}
It is of great interest to try and explain the kinematic behavior seen in a large number of the CJF (combining the percentages calculated from both apparent width and apparent width evolution we get an approximately 30\% of the sample showing wide jets that change their width strongly) and the apparent preference for BL Lacs to show this behavior over FSRQs. One obvious factor that should play a deciding role for the kinematics of jet components is the viewing angle under which a source is observed. It is known that BL Lacs and FSRQs are seen jet-on, at the smallest viewing angles, with steeper-spectrum quasars and radio galaxies being sources observed at progressively larger viewing angles (e.g., \citealt{Antonucci1993}; \citealt{Urry1995}). Could a viewing angle difference explain the kinematic differences seen in some of the CJF sources and in particular observed between BL Lacs and FSRQs? Assuming that jet components follow ballistic, linear paths, one can expect that viewed at smaller viewing angles (smaller than the critical angle $1/\gamma$, with $\gamma$ being the Lorentz factor of the flow), the components are observed to cover smaller distances, than if seen edge-on. Therefore, the slower speeds, and hence smaller total linear evolution of their jet ridge lines, observed for BL Lacs could be explained in terms of a systematically smaller viewing angle for BL Lacs, compared to FSRQs.

However, \citet{Hovatta2009} (based on an investigation initiated by \citealt{Laehteenmaeki1999}) used long timescale variability of AGN, together with jet kinematics, to derive Doppler factors and consequently viewing angles of a sample of 87 AGN. They found that FSRQs actually show smaller mean viewing angles than BL Lacs (a result also found by \citealt{Laehteenmaeki1999}). Although there are certain limitations to the method used by the authors, their results are difficult to reconcile with a viewing-angle-dependent explanation of our results. 

Our results would therefore indicate the need for an additional, potentially geometric, effect in play. This would be in agreement with \citet{Cohen2007}, who postulate that low-speed components in FSRQ and BL Lac jets appear so because their pattern Lorentz factor is lower than the bulk Lorentz factor of the jet. Alternatively, it should be considered whether there is some systematic bias in the way Doppler factors are estimated, that would lead to an over- or underestimation of the viewing angles of BL Lacs and FSRQs, respectively.

The large jet widths and jet widths changes are more difficult to explain. An additional mechanism or effect must be introduced to produce a wide jet. Precession of the jet axis, or assuming that the components follow non-ballistic and non-linear paths can lead to such jet properties (e.g., \citealt{Steffen1995}; \citealt{Gong2008}; \citealt{Roland2008}; \citealt{Gong2011}). In these models, jet components follow highly curved trajectories, which, viewed face-on, would give the impression of a wider distribution of jet component position angles. A precessing jet, or rather a precessing jet nozzle, would imply that different components follow different trajectories. That would result in a changing jet component position angle distribution, more pronounced at smaller viewing angles. That more BL Lacs show M.I. closer to one compared to FSRQs, implying a highly curved, sinusoid-like ridge-line, lends additional support to this scenario. Following this train of thought would then lead us to the conclusion that either (1) there is a sub-set of FSRQs and BL Lacs ($\sim30\%$ in our sample) for which non-ballistic and/or precession effects play an important role, or more generally that (2) for these sources the viewing angles are below some critical angle $
\theta_{c}$ that allow us to witness and thus characterize the helical or ``non-ballistic'' structure common in all radio-AGN jets, but which would otherwise be blended out by the projection effects at viewing angles $>\theta_{c}$.

In a following paper, we shall use an expanded version of the helical jet model of \citet{Steffen1995} to evaluate the statistical results presented here, in terms of the viewing angle (and hence beaming) effect, as well as a possible helical geometry of the jet.

A final scenario that needs to be discussed is whether a systematic difference in the Lorentz factors $\gamma$ between FSRQs and BL Lacs could produce the effects observed here. It has been argued that BL Lacs show smaller Lorentz factors than FSRQs (e.g., \citealt{Morganti1995}; \citealt{Urry1995}; \citealt{Hovatta2009}). If this is true (and not a selection bias in the samples used) then that would naturally explain the slower components in BL Lacs. 
The above in turn ties in to the currently accepted unification scheme (e.g., \citealt{Urry1995}). In that paradigm, BL Lacs and FSRQs are drawn from two, presumably different, parent samples of Fanaroff-Riley I (low luminosity) and II (high luminosity) galaxies (FR; \citealt{Fanaroff1974}; \citealt{Padovani1991}; \citealt{Capetti1999}; \citealt{Xu2009}), respectively. If that is true, then the differences seen in the jet ridge line properties of BL Lacs and FSRQs should then translate to differences between FRI and FRII jet kinematics. Interestingly, studies of FRI and FRII kinematics have shown that both types have similar parcec-scale jet Lorentz factors (e.g., \citealt{Giovannini2001}), contradictory to what we find here, as well as in other studies mentioned above, concerning the parsec-scale jet speeds in BL Lacs and FSRQs. 

It should be noted however that FRI sources show disrupted jets much earlier (i.e., closer to the core) than FRIIs (e.g., see \citealt{Laing1996} and references therein). \citet{Perucho2010} explain this in terms of decollimation through the growth of non-linear Kelvin-Helmholtz instability modes in FRI jets. In this context, the wider jets seen in BL Lacs could indicate larger amplitude instabilities (closer to the non-linear regime) in the jet that, given the lower-speed flows observed in these objects, would lead to an earlier disruption of the jet at kpc scales (as expected for FRIs).

It is difficult to offer a single answer concerning an interpretation of our results. It is very likely that both viewing angle and Lorentz factor effects, as well as flow instabilities that depend on the latter, influence the appearance of BL Lac and FSRQ jets. A comparison between the jet ridge lines of FRIs and FRIIs to BL Lacs and FSRQs, respectively, would offer further insight on the problem. A much larger sample than the CJF would be required for such a study.

\subsection{Comparison with previous studies}
It should be noted that this is the first time the jet ridge lines of a sample of AGN this size have been explicitly and exclusively studied in a statistical manner. While the ridge line of a jet is not a new concept, so far only the jet ridge lines of individual sources (both galactic and stellar) have been studied (e.g., \citealt{Condon1984}; \citealt{Steffen1995}; \citealt{Lister2003};  \citealt{Lobanov2006}; \citealt{Britzen2009}; \citealt{Perucho2009}; \citealt{Lister2009c}; \citealt{Britzen2010}). The definition of the jet ridge line can differ from study to study. We define the ridge line of a jet at a certain epoch as the line that linearly connects the projected positions of all components at that epoch. Alternatively, one can use an algorithm to more directly extract the jet ridge line from the VLBI map of each epoch (e.g., \citealt{Pushkarev2009}). One should however be careful of over-sampling any given map, in effect extracting information that is not actually there. In this sense, the method followed here, although perhaps simpler, is more robust in the quality of information used.

The one obvious direct comparison that can be drawn is to the MOJAVE / 2cm sample (\citealt{Lister2005}; Paper I of a series of 6 papers). As the work presented here is prototype in its conception, it is difficult to draw direct comparisons to the MOJAVE sample. It is of interest to compare the apparent velocity distributions for the MOJAVE and the CJF, as the apparent velocities are in a sense reflected in our measure of the total jet ridge line linear evolution. The CJF sources appear to be significantly slower than what is seen in the MOJAVE sample (e.g., compare Fig. 7 from \citealt{Britzen2008} and Fig. 7 from \citealt{Lister2009c}). In the distribution of apparent speeds, for the CJF there is a maximum at around 4c and then a turn down, with very few sources above 15c. In contrast, the distribution for the MOJAVE sources shows a maximum at 10c with a fair number of sources showing speeds greater than 20c. This can be explained in terms of the higher temporal resolution of the MOJAVE sample (owning to the larger number of epochs per source) that might allow the detection of such faster motions. It is interesting to note that no clear distinction has been made in the MOJAVE sample between BL Lacs and FSRQs, probably due to the relatively small number of BL Lacs. \citet{Lister2009c} however do mention that BL Lacs are more often found to exhibit ``low-pattern speeds", i.e., components with slow motions, significantly smaller than others in the same jet.  Similar evidence has been previously found in different VLBI samples (e.g., \citealt{Jorstad2001}; \citealt{Kellermann2004}). Combined with the argument that BL Lac jets exhibit lower Lorentz factors, as a result of a possible correlation between intrinsic AGN luminosity and jet Lorentz factor (e.g., \citealt{Morganti1995}), the above are in agreement with CJF BL Lacs showing on average weaker linear evolution than FSRQs. It should be noted that the agreement with previous studies concerning the outward motions of BL Lacs objects supports the robustness and reliability of our method.

As was previously mentioned, a correlation between apparent speeds and luminosity has been studied for the MOJAVE sample too (\citealt{Cohen2007}; \citealt{Lister2009c}). A similar correlation was observed also for the CJF sample, as was shown in \citet{Britzen2008}. It should be noted that for both samples, these correlations were between luminosity and either the maximum apparent speed per source (in the case of MOJAVE), or only for the best quality (Q1) data (for the CJF). In this paper we presented a similar correlation between 5 GHz core luminosity (as derived from VLA measurements) and the total linear evolution of the jet ridge line. In this sense, we include speed information for all components identified across-epochs and therefore treat the jet as a whole, rather than singling out the fastest component. The result is however remarkably similar to what was previously found. \citet{Cohen2007} and \citet{Lister2009c} discuss possible interpretations of the upper-envelope that is observed in this correlation, in terms of the beaming model. They conclude that there is no apparent argument as to why such an envelope should exist. They instead offer an alternative interpretation in terms of a link between the energy output of the AGN and the intrinsic Lorentz factor of a jet. They conclude however that the shallow flux limit of the MOJAVE sample does not allow for a robust conclusion. The deeper flux-limit of the CJF (and the resulting more than double number of sources) allow us to re-address the same issue. The fact that this correlation remains in the CJF sample, both in the sense of a maximum apparent jet speed, as well as that of the linear evolution measure (in essence a mean jet speed), argues in favor of a link between the intrinsic luminosity and Lorentz factor. A definite answer would come by the inclusion of low-luminosity sources in such an investigation. It would then become obvious whether this envelope holds and potentially differentiate between a beaming model signature and that of an intrinsic connection of the energy output and the jet kinematics.

A final note concerning the comparison between our results and these of the MOJAVE sample pertains to the observations frequency of each sample. Unlike the CJF, the MOJAVE uses VLBI observations at 15 GHz, meaning that a potentially different regime of the jet is probed, compared to what is studied in the CJF. Two-zone models, such as the ``two-fluid'' model (e.g., \citealt{Sol1989}, \citealt{Pelletier1989}), predict a fast jet spine, embedded in a slower moving sheath. This would imply that the MOJAVE potentially probes deeper, thus faster, jet layers compared to what is seen in the CJF sample. Moreover, higher observation frequency also translates to higher angular resolution and therefore to probing effects closer to the core and also enabling the detection of faster moving components. In this case, it would be interesting to apply the methods used in this paper for the MOJAVE dataset, as it is expected that the curvature of the jet (and therefore potentially the angular properties and evolution of the jet ridge line) should play an increasingly important role at closer core separations.

\section{Conclusions}

We summarize our results:
\begin{itemize}
    \item we develop a number of tools to investigate both the morphology and kinematics of the CJF jet ridge lines:
        \begin{enumerate}
            \item Monotonicity Index
            \item Apparent jet width
            \item Apparent jet width evolution
            \item Apparent jet linear evolution
        \end{enumerate}
    \item Using the M.I., we find that BL Lacs jet ridge lines more often resemble a sinusoidal curve compared to both FSRQs and RGs. In contrast, 2/3 of the FSRQs have an M.I. lower than 0.5 (indicating fairly monotonic jets).
    \item 22.4\% of the CJF sources have apparent jet widths larger than 20 degrees. 47.3\% of BL Lacs, over 13.6\% and 25\% for FSRQs and RGs, respectively, have $dP>20$ degrees.
    \item BL Lacs exhibit substantially apparently wider jets. This effect persists under several constraints (e.g., redshift bins, core separation limits). Quasars show the least wide jets among the three classes. This supports the effects seen in individual BL Lac objects (i.e., 1803+784, 0716+714, etc.).
    \item No indication is found for a change of apparent jet width with redshift, at least out to $z\sim2$.
    \item The distribution of apparent jet ridge line widths for BL Lacs appears to extend towards higher values, with FSRQs mainly contained at lower values. Radio galaxies show a somewhat wider distribution of widths than FSRQs. A K-S test indicates a $0.7\cdot10^{-3}\%$ probability that FSRQs and BL Lacs are drawn from a single parent population.
    \item 28.3\% of the CJF sources exhibit apparent jet width evolution larger than 4 deg/yr. 12.5\% of BL Lacs have $\Delta P_{max}>4$, as compared to 4\% and 5.7\% for FSRQs and RGs, respectively.
    \item When taking into account the maximum values of apparent width evolution for the CJF sources, BL Lacs are found to show marginally stronger evolution of their apparent jet widths, with FSRQs and RGs showing similar values (within errors). This behavior holds for a smaller redshift bin.
    \item BL Lac objects, on average, show weaker apparent linear evolution of their jet ridge lines compared to FSRQs. Interestingly, RGs show even weaker linear evolution. BL Lac objects and radio galaxies show apparent slower moving jets than quasars.
    \item Variable CJF source are, on average, found to exhibit apparent wider jets than the non-variable ones. A K-S test gives a 1\% probability that these two sub-samples stem from the same parent distribution.
    \item Variable sources in the CJF show, on average, stronger apparent angular evolution than the non-variable ones. This is reflected in the fairly wider distribution of apparent jet width evolution angles compared to the parent (non-variable) sample.
    \item Variable sources exhibit stronger apparent linear evolution. This is seen also in the distributions of the two sub-samples.
    \item Variable FSRQs show more ``BL Lac-like" behavior with wider jets that also change their width more strongly.
   \item A correlation, in the form of an upper envelope, is found between total apparent linear evolution of the CJF jet ridge lines and their radio (VLA, 5 GHz) core luminosities. Although partly attributed to relativistic effects, there are indications of a further, intrinsic effect additional to the geometric one.
\end{itemize}

In conclusion, by statistically analyzing the CJF sample, we provide detailed insight concerning the morphology and evolution of AGN jet ridge lines. The statistical investigation of the CJF sources lends independent support to the different kinematic scenario recently seen in a number of BL Lac objects (1803+784, \citealt{Britzen2010}; 0735+178, \citealt{Britzen2010b}; etc.). 25\%-30\% of the CJF sample show apparently considerably wide jets and strong apparent width evolution. BL Lac objects appear to deviate the strongest from the kinematic paradigm, widely accepted for blazars, of outward superluminally moving jet components. BL Lacs appear to evolve their jet ridge lines (with respect to the core) less than the other source classes, hence indicating a slower apparent flow in their jets. On the other hand, they show significantly apparent wider jets, with a trend to change their widths more strongly, than both FSRQs and RGs. All these three effects have been already observed in the jets of 1803+784, 0716+714, and 0735+178. Viewing BL Lacs at small angles to their jet axis possibly allows us to uncover this peculiar kinematic behavior that would otherwise be inaccessible at larger viewing angles. However, the picture does not remain ``simple". By studying a sub-sample of CJF sources identified as variable, we find that these sources actually show wider, more strongly evolving jet ridge lines, with variable FSRQs exhibiting a more ``BL Lac like" behavior. All the above, combined with the significant number of CJF sources showing evidence supporting a different kinematic scheme (independent of their classification), imply a rather universal effect, rather than something unique to BL Lacs.

It should therefore be noted that the above results underline the fact that the notion of linear, ballistic trajectories for AGN jet components usually employed until recently is a very crude approximation and, more often than not, deviates grossly from the reality. It is of great interest to uncover the process that leads to the properties of AGN jet ridge lines as studied in this paper. In a following paper, we shall use a simple jet model to address this point.

\begin{acknowledgements}
M. Karouzos was supported for this research through a stipend from the International Max Planck Research School (IMPRS) for Astronomy and Astrophysics. M.K. wants to thank Lars Fuhrmann and Mar Mezcua, as well as an anonymous referee for insightful discussions and comments that greatly improved this manuscript. This research has made use of the NASA/IPAC Extragalactic Database (NED) which is operated by the Jet Propulsion Laboratory, California Institute of Technology, under contract with the National Aeronautics and Space Administration. This research has made use of NASA's Astrophysics Data System Bibliographic Services. The Very Long Baseline Array is operated by the USA National Radio Astronomy Observatory, which is a facility of the USA National Science Foundation operated under cooperative agreement by Associated Universities, Inc.
\end{acknowledgements}

\bibliographystyle{aa}
\bibliography{bibtex}




\end{document}